\documentclass[aps,pre,notitlepage]{revtex4-1}
\usepackage{amssymb,amsmath}
\usepackage{amsthm}
\usepackage{graphicx}
\usepackage[caption=false]{subfig}
\usepackage{float}
\usepackage{color}
\usepackage{mathscinet}

\begin{document}

\title{Ordering dynamics in the voter model with aging}
\author{Antonio F. Peralta}
\author{Nagi Khalil}
\author{Ra\'ul Toral}
\affiliation{IFISC (CSIC-UIB), Instituto de F\'isica Interdisciplinar y Sistemas Complejos, Campus Universitat de les Illes Balears, E-07122 Palma de Mallorca, Spain}

\begin{abstract}
The voter model with memory-dependent dynamics is theoretically and numerically studied at the mean-field level. The ``internal age'', or time an individual spends holding the same state, is added to the set of binary states of the population, such that the probability of changing state (or activation probability $p_i$) depends on this age. A closed set of integro-differential equations describing the time evolution of the fraction of individuals with a given state and age is derived, and from it analytical results are obtained characterizing the behavior of the system close to the absorbing states. In general, different age-dependent activation probabilities have different effects on the dynamics. When the activation probability $p_i$ is an increasing function of the age $i$, the system reaches a steady state with coexistence of opinions. In the case of aging, with $p_i$ being a decreasing function, either the system reaches consensus or it gets trapped in a frozen state, depending on the value of $p_\infty$ (zero or not) and the velocity of $p_i$ approaching $p_\infty$. Moreover, when the system reaches consensus, the time ordering of the system can be exponential ($p_\infty>0$) or power-law like ($p_\infty=0$). Exact conditions for having one or another behavior, together with the equations and explicit expressions for the exponents, are provided.
\end{abstract}
\maketitle

\section{Introduction}\label{intro}

Agent-based binary-state models are commonly used to study the effect that simple individual dynamical rules may have on the collective behavior of a system. Typical examples include modeling the spreading of diseases \cite{Disease1,Disease2}, the time evolution of the number of speakers of a determined language \cite{Language1,Language2,Language3,Language4}, the evolution of prices in financial markets \cite{Kirman,Markets1,Markets2,Markets3,Markets4,Carro1,KONONOVICIUS2019171}, or the dynamics of opinion formation in a population of individuals \cite{babave08,cafolo09,PhysRevLett.112.158701}. As a prominent case of the latter, the voter model \cite{clsu73,holi75,Vazquez_2008} and its variants thereof \cite{2018arXiv181111888R}, by which an agent adopts with some probability the state of a randomly chosen neighbor, are widely used when the mechanism at hand is that of imitation.  Beyond its potential applications, the voter model has become a paradigm in the field of non-equilibrium statistical physics. It is defined by very simple rules, and explicit solutions can be given in some particular cases, yet it is capable of showing very rich behavior, including phase transitions, characterized by critical exponents, power-law correlations, finite-size scaling, tricritical behavior, etc. \cite{krrebe10,li12}. Recent studies of the voter model include the effect of a network structure \cite{Peralta_pair,Carro2}, non-linear or group interactions \cite{Nyczka2012,Nyczka2013,Jedrzejewski2017,Peralta,PhysRevE.97.052106,Raducha_2018,Min_2019}, more than two states \cite{herrerias:2019,vazquez:2019}, the effect of zealots \cite{Nagi} and contrarians \cite{Nagi2}, etc.

A fundamental aspect of the agent-based binary-state models is the memory the constituents of the system have \cite{JEDRZEJEWSKI2018306}, and how this affects the dynamical rules that define the model. A clear example is the framework of non-Poissonian infection processes in epidemic spreading \cite{Min1,Boguna2014,Gleeson}, where the infection of an individual depends on the times each one of its neighbors became infected \cite{Blythe}. This is because the rate at which one transmits a disease is not independent of the time one has been infected. In most cases of binary-state models a Markovian assumption is postulated which implies that the probability of changing state is independent of the past history of events of the individuals, that is to say neglecting memory effects by including Poissonian processes. This is of course a simplification that can be justified in many phenomena and it makes a lot easier the mathematical treatment \cite{nonMarkov1,nonMarkov2}, but it may be lacking of realistic features in other cases. In this context, one of the objectives of the present work is to provide a consistent mathematical description of the voter model including memory effects, namely by making the probability of changing opinion of agents to depend on their age or time elapsed until their last update \cite{Juan,Schweitzer}. 

One of the main questions addressed by the voter model is whether the imitation rule leads to a situation of consensus, with all agents (or at least, a significant majority) adopting the same value of the binary state \cite{befrkr96,sore05,suegsa05}. The answer to this question is sometimes counter-intuitive. Although it would appear, for example, that an increase in the connections amongst the agents should favor consensus, a detailed analysis shows that for an effective dimensionality of the connectivity network greater than two, the system is not able to sustain a consensus state in the thermodynamic limit. The problem has been also addressed when including memory effects. Stark et. al \cite{Schweitzer} introduced increasing inertia, otherwise known as ``aging'', as a mechanism that reduces the probability of an agent to copy the state of another one. Aging can roughly be related to an increase of the inflexibility in the copying mechanism \cite{martins:2013}. Again, it would seem naively that reducing the number of interactions amongst agents should impede the reaching of consensus, but it was shown that exactly the opposite happens, namely the consensus is reached more easily in such an scenario of aging. In a recent paper by some of us \cite{Oriol,2018arXiv181205378A}, it was shown that the inclusion of aging in the noisy version of the voter model \cite{PhysRevLett.63.2857,Kirman,Granovsky} induces a state of imperfect consensus that remains in the thermodynamic limit. Another study of the noiseless voter model \cite{Juan} focused on the distribution of the time between individual changes of states $C(t)$, which in turn is related with the activation probability $p_i$, or the age-dependent probability of interaction. The conclusion of \cite{Juan}, sustained from the analysis of extensive numerical simulations, was that a particular functional form of $p_i \sim \beta/i$ induces a power-law dependence $C(t)\sim t^{-\beta}$, which has been observed in real datasets, for example in human communications \cite{Wu,Candia}. This power law is similar to the ones observed in the approach to consensus in low-dimensional lattices, but it shows a strong dependence on the details of the activation probability. The main objective of this paper is to analyze this problem and to derive the relations in \cite{Juan,Schweitzer}, among others, from an analytical point of view. We also aim to give a well constructed theoretical framework for aging, which has been studied together with network structure in several recent works \cite{Perez,Oriol3,Oriol2} but mainly computationally, due to difficulties encountered in the mathematical description.

The outline of the paper is as follows. In section \ref{model_def} we introduce the voter model including memory effects. Two specific forms of the activation probability, rational and exponential functions of the age, are considered which depend on several parameters, so as to cover phenomenology observed in the literature. Section \ref{absorbing_states} includes a summary of the main results present in the literature about dynamical behavior of the system close to the absorbing states. In section \ref{sec:master} the dynamical equations of the mean fraction of agents with a given state and age are obtained under a mean field description and its steady-state analysed. Section \ref{noneq} contains the main results of the work. First, a closed set of integro-differential equations describing the system close to the absorbing states is derived. Second, an approximate analysis of the previous description is carried out, providing explanation to the numerical findings. Complementary results are given in Appendix \ref{app1} and Appendix \ref{appb}. Finally, section \ref{summary_conclusions} includes a summary of the main results.

\section{Model}
\label{model_def}

The voter model implements in arguably the simplest way the herding mechanism for the evolution of binary state systems. In the original version one considers a system formed by $N$ individuals or agents. Agents are located in the nodes of a network and are connected by undirected, bidirectional, links. Two connected individuals are said to be ``neighbors". The network is single connected, meaning that every node can be reached from any other node by a sequence of links. Each individual $k=1,\dots,N$ holds a binary-state (spin) variable $s_k=\pm1$. Several interpretations can be given to this binary variable, but its exact meaning does not concern us in this paper. For example, the voter model or variants of it have been used to represent the optimistic/pessimistic state of a stock market broker \cite{Kirman}, the language A/B used by a speaker \cite{Language1,Language4}, or the direction of the velocity right/left in a one-dimensional model of active particles \cite{Escaff:2018}. Those variables evolve over time by the following (stochastic) rules:\\[5pt]
(i) An individual $k$ is selected at random amongst the $N$ possibilities.\\
(ii) The selected individual copies the state $s_k=s_{k'}$ of another individual $k'$ chosen also at random between the set of neighbors of $k$. \\[5pt]
In this sequential updating scheme, every time an individual is chosen for updating, time $t$ increases by one unit, while $N$ updates constitute one Monte Carlo step (MCS).

In the aging version of the voter model the above rules are modified such that the herding mechanism occurs only with an \emph{activation probability} $p_{i_k}$ that depends on the {\slshape internal age} $i_k$ of the selected individual. The internal age $i_k=0,1,2,\dots$ of individual $k$ stands for the number of update attempts elapsed since its last change of state. 

More explicitly, the model is reproduced in the simulations by modifying the last step as follows:\\[5pt]
(ii) The selected individual copies {\bf with probability} $\mathbf{p_{i_k}}$ the state $s_k=s_{k'}$ of another individual $k'$ chosen at random between the set of neighbors of $k$. If the selected individual changes state $s_k \rightarrow -s_k$, then its internal age resets to zero $i_k \rightarrow 0$; otherwise it increases in one unit $i_k \rightarrow i_k+1$. Initially, all internal ages are set to zero.

The standard (no aging) voter model is recovered taking $p_i=1$. Here, for the sake of concreteness and mathematical treatment, we have considered the following two functional forms for the activation probability:
\begin{itemize}
\item[1.-] The first functional form is a rational function of the age,
\begin{equation}
\label{eq:pv1}
p_i=\frac{p_\infty i+p_0c}{i+c},\quad i=0,1,2,\dots
\end{equation} 
where $p_0,\,p_\infty\in[0,1]$ and $c>0$ are constants. If $p_0>p_\infty$, the probability of interaction decreases with age, a typical aging situation. The opposite occurs if $p_0<p_\infty$, as the activation probability increases with age. This can be interpreted as ``anti-aging'' or ``rejuvenating'' where as nodes get older they become more prone to change state. This aging form, with $p_\infty=0$, is basically the form considered in \cite{Juan} and that has been shown to induce features observed in several real-word systems, such as power-law inter-event time distributions. If $p_0=1$ and $p_\infty>0$, this reproduces qualitatively the form used in Ref. \cite{Schweitzer}. In that paper, the activation probability $p_i$, equivalent to $1-\nu_i$ with $\nu_i$ being the ``inertia'', decreases linearly from $p_0=1$ up to a final constant, non-null value $p_\infty$, but at a finite value of $i$.
\item[2.-] The second form we adopt for mathematical simplicity is an exponential dependence of the age,
\begin{equation}
\label{eq:pv2}
p_i=(p_0-p_\infty)\lambda^i+p_\infty,\quad i=0,1,2,\dots
\end{equation} 
with the same meaning as before for $p_0$ and $p_\infty$, and $\lambda<1$ is a parameter.

\end{itemize}

\section{Dynamical behavior. Absorbing states}\label{absorbing_states}

When one individual changes its state after copying the state of one of its neighbors, it might occur that it decreases the total number of neighbors with which it shares the same state. Therefore, a question of interest is whether by iteration of the dynamical rules a ``consensus'', or fully ordered, state in which all agents share the same state, either $+1$ or $-1$, is finally achieved. One can argue that the stochastic rules permit eventually to reach any possible configuration of states. Therefore, when any of the particular consensus states $s_k=+1,\forall k$ or $s_k=-1,\forall k$ is reached, then there can not be further evolution and the dynamics stops. It seems, though, that the ultimate fate of the system is to reach consensus, with all nodes sharing the same value. While this is certainly true for any system with a finite number $N$ of agents, it is interesting to analyze the way the consensus state is reached and the dependence with system size of the average time to reach that state. Two different measures have been introduced to study the approach to order: the magnetization $m(t)=N^{-1}\sum_{k=1}^Ns_k(t)$ that takes values $m=\pm 1$ in the ordered states, and the density $\rho(t)$ of active links, defined as the fraction of links that join nodes in different states. In the ordered states, the magnetization can be $m=+1$ or $m=-1$, while $\rho=0$ in both absorbing states. The existence of absorbing states then ensures that $\lim_{t\to\infty}m(t)=\pm 1$ and $\lim_{t\to\infty}\rho(t)=0$ for a finite system.

There is a significant difference between the aging and non-aging versions of the model. In the non-aging version the approach to the absorbing state depends crucially on the spatial dimension $d$. If $d> 2$ it is observed that the density of active links \footnote{A similar qualitative behavior is observed for the magnetization. In fact, for the all-to-all connectivity considered later in this paper, one has simply $\rho(t)=\frac12\left[1-m(t)^2\right]$, although in other networks $\rho(t)$ must be considered as an independent variable. } has a fast transient to a plateau value $\rho^{*}$ (for $d \rightarrow\infty$, i.e. all to all connectivity, $\rho^{*}=\rho(0)$) and it oscillates around it until a large fluctuation brings the system to the absorbing state $\rho=0$ at a time $T$. This time is a stochastic variable whose mean value scales as $\langle T\rangle \sim N$ for $d>2$, see also \cite{Nagi3} where the probability distribution of $T$ is obtained for $d\to \infty$. Therefore, larger systems stay longer in an {\sl active} state until a large fluctuation brings them to the absorbing state. As we take the thermodynamic limit $N\to\infty$, it is $T\to\infty$ and the system does not order at all in a finite time. If we average over initial conditions and realizations of the dynamics, we find an exponential decay $\langle \rho(t)\rangle=\rho^{*} e^{-t/\tau}$, with $\tau \propto \langle T\rangle$. As mentioned, in the thermodynamic limit when $\langle T\rangle=\infty$, we obtain $\langle \rho(t)\rangle=\rho^{*}$, or absence of order. If we take first the limit $t\to \infty$ and then the limit $N\to\infty$ we obtain otherwise $\lim_{N\to\infty}\lim_{t\to\infty}\langle \rho(t)\rangle=0$. If the spatial dimension is $d\le 2$ the decay to the absorbing state occurs very differently. One does not observe that $\rho(t)$ fluctuates around a plateau value, but rather $\rho(t)$ decreases on average for any realization even for large system sizes. As a consequence, in the thermodynamic limit, it is predicted that $\lim_{t\to\infty}\lim_{N\to\infty}\langle \rho(t)\rangle=0$. The exact time dependence depends on the dimension as $\lim_{N\to\infty}\langle \rho(t)\rangle=(\log t)^{-1}$ for $d=2$ and $\lim_{N\to\infty}\langle \rho(t)\rangle=t^{-1/2}$ for $d=1$ \cite{befrkr96,sore05,suegsa05}. 

The phenomenology in the aging version was described in \cite{Juan} for particular forms of the activation probability. It was observed that if $p_{i}$ follows a similar dependence as given by Eq.  (\ref{eq:pv1}) with $p_\infty=0$, then the approach to the absorbing states is described by a power law $\lim_{N\to\infty}\langle \rho(t)\rangle=t^{-\beta}$, for large spatial dimensions. Furthermore, in a fully-connected network it was observed numerically that the exponent $\beta$ of the power-law decay agreed within the numerical accuracy with the parameter $b = p_{0} c$ of Eq.  (\ref{eq:pv1}) \footnote{Strictly speaking, the aforementioned power-law dependence was reported in \cite{Juan} for the cumulative inter-event time distribution $C(t)$. One of the authors of that reference (J. F. Gracia, private communication) has confirmed to us that the aforementioned asymptotic time dependence of $\rho(t)$ was also observed, but not displayed.}. This indicates that in the presence of aging, the system always orders, contrarily to the non-aging version. In this paper we offer an explanation of this behavior. 

\section{The dynamical equations}
\label{sec:master}
We now derive the mean-field dynamical equations of our stochastic process, as defined in section \ref{model_def}. In the derivation we restrict ourselves to the all-to-all (or fully connected) network in which all nodes are neighbors. The mathematical description in the case of a more complex network structure in the interactions between nodes is a further complication \cite{Peralta_pair} and it is left for future studies \cite{Oriol3,Oriol2}. In the all-to-all setup, all the information needed to implement the stochastic update rules is contained in the set $S\equiv \lbrace n^{\pm}_{i} \rbrace_{i=0}^\infty$ of the numbers of individuals with internal age $i$ in states $\pm 1$, \cite{Schweitzer}. The global variables for the total number of up and down spins are $n = \sum_{i=0}^{\infty} n_{i}^{+}$ and $N-n= \sum_{i=0}^{\infty} n_{i}^{-}$. Note, however, that not all variables of the state $S$ are independent, since $\sum_{i=0}^\infty (n_i^++n_i^-)=N$. Hence, it is useful to consider an alternative representation of the system in terms of independent variables, for instance by using the variable $n$ and obviating $n_0^\pm$, as $S_n\equiv \left(n,\lbrace n_i^+\rbrace_{i=1}^\infty,\lbrace n_i^-\rbrace_{i=1}^\infty\right)= \left(n,n_1^+,n_2^+,\dots,n_1^-, n_2^-, \dots\right)$. In this representation $n_0^\pm$ are given by $n_0^{+}=n-\sum_{i=1}^\infty n_i^+$ and $n_0^{-}=N-n-\sum_{i=1}^\infty n_i^-$.

In order to derive the dynamical equations of the global state of the system $S_n$ we must include with their respective probabilities all events that induce changes in $S_n$. Every time an individual changes its state or age, there is a change in the set of variables $S_n$. The change can occur in different ways and with different probabilities. We now spell in detail the possibilities and their effects on the variables.\\

{\bf(1)} Consider that at time $t$ the chosen individual $k$ is in state $s_k=+1$ and has age $i_k=i$ and that as a result of the interaction it switches to $s_k=-1$ and, hence, its internal age is reset to $i_k=0$. The probability of this event is equal to the probability, $\tfrac{n_i^+}{N}$, of choosing and individual with age $i$ and state $+$, multiplied by the age-dependent probability, $p_i$, that it activates the copying (herding) mechanism, and multiplied by the probability $\tfrac{N-n}{N}$ that the randomly selected neighbor is in the opposite state. Altogether, the probability is $\tfrac{n_i^+}{N} p_i \tfrac{N-n}{N}\equiv \tfrac{1}{N}\Omega_{1,i}$. When the switching of state occurs, if $i>0$ we have $n_{i}^{+} \rightarrow n_{i}^{+}-1$ and $n \rightarrow n-1$, while if $i=0$ the only change is $n\to n-1$.\\

{\bf(2)} This is similar to the previous case but now the chosen individual is initially in state $s_k=-1$. The probability of switching is $\tfrac{n_i^-}{N} p_i \tfrac{n}{N}\equiv \tfrac{1}{N}\Omega_{2,i}$. When the switching of state occurs, if $i>0$ we have $n_{i}^{-} \rightarrow n_{i}^{-}-1$ and $n \rightarrow n+1$, while if $i=0$ the only change is $n\to n+1$.\\

{\bf(3)} Consider that at time $t$ the chosen individual $k$ has $s_k=+1$ and age $i_k=i$, but that now it keeps its current state $s_k=+1$. This event happens with a probability equal to the probability of choosing an individual in state $+1$ with age $i$, $\tfrac{n_i^+}{N}$, multiplied by the probability that it does not switch, which can arise either because the copying mechanism was not activated, with probability $1-p_i$, or it was activated but the selected neighbor was in the same state, with probability $p_i\tfrac{n_i^+}{N}$. Altogether, the probability is $\tfrac{n_i^+}{N} \left(1-p_i+p_{i}\tfrac{n}{N} \right) \equiv \tfrac{1}{N}\Omega_{3,i}$. In this case the variables change as $n_{i}^{+} \rightarrow n_{i}^{+}-1$ and $n_{i+1}^{+} \rightarrow n_{i+1}^{+}+1$ if $i>0$ and $n_{1}^{+} \rightarrow n_{1}^{+} +1$ if $i=0$.\\

{\bf(4)} Finally, we consider a similar case to the previous one but the chosen individual $s_k=-1$ keeps its state. The switching probability is now $\tfrac{n_i^-}{N}\left(1-p_i+p_i\tfrac{N-n}{N}\right)\equiv\tfrac{1}{N}\Omega_{4,i}$. The changes in the state $S_n$ are $n_i^-\to n_i^--1$, $n_{i+1}^-\to n_{i+1}^-+1$ if $i>0$ and $n_1^-\to n_1^-+1$ if $i=0$.\\

If time is measured in MCS, the rates (probability per unit time) of the four possible processes are
\begin{eqnarray}
\Omega_{1,i}&=&n_i^+\beta_i(1-n/N), \quad\Omega_{2,i}=n_i^-\beta_i(n/N),\\
\Omega_{3,i}&=&n_i^+\alpha_i(1-n/N), \quad\Omega_{4,i}=n_i^-\alpha_i(n/N),
\end{eqnarray}
where $\beta_i(x)=p_i\,x$ and $\alpha_i(x)=1-\beta_i(x)$, thus $\Omega_{1,i}+\Omega_{3,i}=n_i^+$, $\Omega_{2,i}+\Omega_{4,i}=n_i^-$.

This derivation has considered a discrete time process in the limit of a large number of individuals. It is also possible to consider from the beginning a continuous time process in which an agent of age $i$ and state $s=+1$ has a rate $\beta_i(1-n/N)$ of switching state and set its internal age to $0$, and a rate $\alpha_i(1-n/N)$ of increasing its internal age keeping the state; similarly an agent of age $i$ and state $s=-1$ has a rate $\beta_i(n/N)$ of switching state and set its internal age to $0$, and a rate $\alpha_i(n/N)$ of increasing its internal age and keep the state.

Following standard techniques \cite{vanKampen:2007,Toral-Colet:2014,Peralta_moments}, we could derive a master equation for the probability $P(S_{n},t)$ of having state $S_{n}$ at time $t$, with the set of rates and changes in the variables explained before. As we restrict ourself to average values in this paper, we will only obtain the evolution equations for the ensemble average of the fraction of nodes with a given state and age $x_i^\pm =\langle n_i^\pm\rangle/N$, as well as for $x = \langle n \rangle/N$. The time evolution of the average values can be computed as a weighted sum of all the rates that contribute to its variation, the weights being the variation of the regarded variable in that process \cite{Peralta_moments}. Using the standard mean-field approximation \cite{Toral-Colet:2014} which neglects correlations as $\langle n_i^\pm n \rangle \simeq \langle n_i^\pm\rangle \langle n \rangle$ we end up with a closed but infinite system of equations:
\begin{eqnarray}
\label{eq_dyn1}
\frac{d x_{i}^{+}}{dt} &=& -x_{i}^{+} + x_{i-1}^{+} \alpha_{i-1}(1-x), \quad i\ge 1, \\
\label{eq_dyn2}
\frac{d x_{i}^{-}}{dt} &=& -x_{i}^{-} + x_{i-1}^{-} \alpha_{i-1}(x), \quad i\ge 1,\\
\label{eq_dyn3}
\frac{dx}{dt} &=& \sum_{i=0}^{\infty} x_{i}^{-} \beta_{i}(x) - \sum_{i=0}^{\infty} x_{i}^{+} \beta_{i}(1-x).
\end{eqnarray}
Similar equations were obtained in Ref. \cite{Schweitzer,Oriol}. In these equations, variables $x_0^\pm$, whenever they appear, should be expressed in terms of the independent variables,
\begin{eqnarray}
 \label{eq:x01}
 x_0^+=x-\sum_{i=1}^{\infty}x_{i}^{+}, \qquad x_0^- = 1-x - \sum_{i=1}^{\infty} x_{i}^{-}.
\end{eqnarray}
The explicit time evolution of these variables is
\begin{eqnarray}
\label{eq_dyn4}
\frac{d x_{0}^{+}}{dt} &=& -x_{0}^{+} + \sum_{i=0}^{\infty} x_{i}^{-} \beta_{i}(x), \\
\label{eq_dyn5}
\frac{d x_{0}^{-}}{dt} &=& -x_{0}^{-} + \sum_{i=0}^{\infty} x_{i}^{+} \beta_{i}(1-x).
\end{eqnarray}

By equating all time derivatives to zero, we can identify the steady-state solutions for the mean-field description. From Eqs. (\ref{eq_dyn1},\ref{eq_dyn2},\ref{eq:x01}) we find
\begin{eqnarray}
\label{eq:x03}
&& x_{i,\mathrm{st}}^+=\dfrac{x_\mathrm{st}}{f(x_\mathrm{st})}\prod_{k=0}^{i-1}\alpha_{k}(1-x_\mathrm{st}), \hspace{0.5cm} x_{i,\mathrm{st}}^-=\frac{1-x_\mathrm{st}}{f(1-x_\mathrm{st})}\prod_{k=0}^{i-1}\alpha_{k}(x_\mathrm{st}),\quad i\ge 0,
\end{eqnarray}
where
\begin{equation}
\label{eq:funf}
 f(x)\equiv \sum_{i=0}^\infty \prod_{j=0}^{i-1}\alpha_{j}(1-x),
\end{equation}
and the convention $\prod_{j=0}^{-1}\alpha_{j}\equiv 1$.
Obviously, the validity of these relations require that the series Eq. (\ref{eq:funf}) defining $f(x)$ is convergent. This is certainly not the case for $x=1$, as $f(x=1)=\infty$ always. As we will see, another case in which the series might diverge occurs when $p_{\infty} = 0$, when d'Alembert's criterion does not ensure convergence as $\lim_{i\rightarrow \infty} \alpha_{i}(x) = 1$. 
Using Eqs. (\ref{eq_dyn3},\ref{eq_dyn4},\ref{eq_dyn5}) in the steady state, one obtains easily $x_{0,\mathrm{st}}^+=x_{0,\mathrm{st}}^-$, or
\begin{equation}
 \label{eq:xest}
 \frac{x_{\mathrm{st}}}{f(x_{\mathrm{st}})}=\frac{1-x_{\mathrm{st}}}{f(1-x_{\mathrm{st}})}.
\end{equation}
The solutions to this equation provide the possible steady-state values of $x_\text{st}=\langle n\rangle /N$ and those values of $x_\text{st}$ determine the other quantities, through Eqs. (\ref{eq:x03}). Note that $x_\text{st}=1/2$ is always a trivial solution that corresponds to a symmetric steady state, with the same mean number of nodes with a given age having opposite states $x_{i,\mathrm{st}}^+=x_{i,\mathrm{st}}^-$ for $i\ge 0$. In any case, the steady-state solutions describe situations where $x_{i,\mathrm{st}}^\pm$ are decreasing functions of the age. Note also that $x_{\mathrm{st}}=0,1$ are always steady state solutions (absorbing states) of the dynamics, and indeed satisfy Eq. (\ref{eq:xest}) as $f(x \rightarrow 1) \rightarrow \infty$ and $f(x\rightarrow 0) \rightarrow \text{constant}$.

For the particular case $p_i=p,\forall i$, we obtain the function \makebox{$f(x)=1/(p(1-x))$}. Eq. (\ref{eq:xest}) is satisfied for any $x_{\mathrm{st}}$ and, in fact, it can be shown that Eq. (\ref{eq_dyn3}) becomes $\frac{dx}{dt}=0$ and, hence, $x_{\mathrm{st}}=x(0)$. This is the behavior of the non-aging voter model in the thermodynamic limit that was described before. If $x_{\mathrm{st}}\ne 0,1$, the distribution of ages in this steady state follows a geometric distribution: $x_{i,\mathrm{st}}^+=px_{\mathrm{st}}(1-x_{\mathrm{st}})(1-p(1-x_{\mathrm{st}}))^i$ and $x_{i,\mathrm{st}}^-=px_{\mathrm{st}}(1-x_{\mathrm{st}})(1-px_{\mathrm{st}})^i$. For $x_{st}=0,1$, see section \ref{noneq}.

\begin{figure}[h!] 
\centering
\subfloat[]{\label{fig:fixeda}\includegraphics[width=0.45\textwidth]{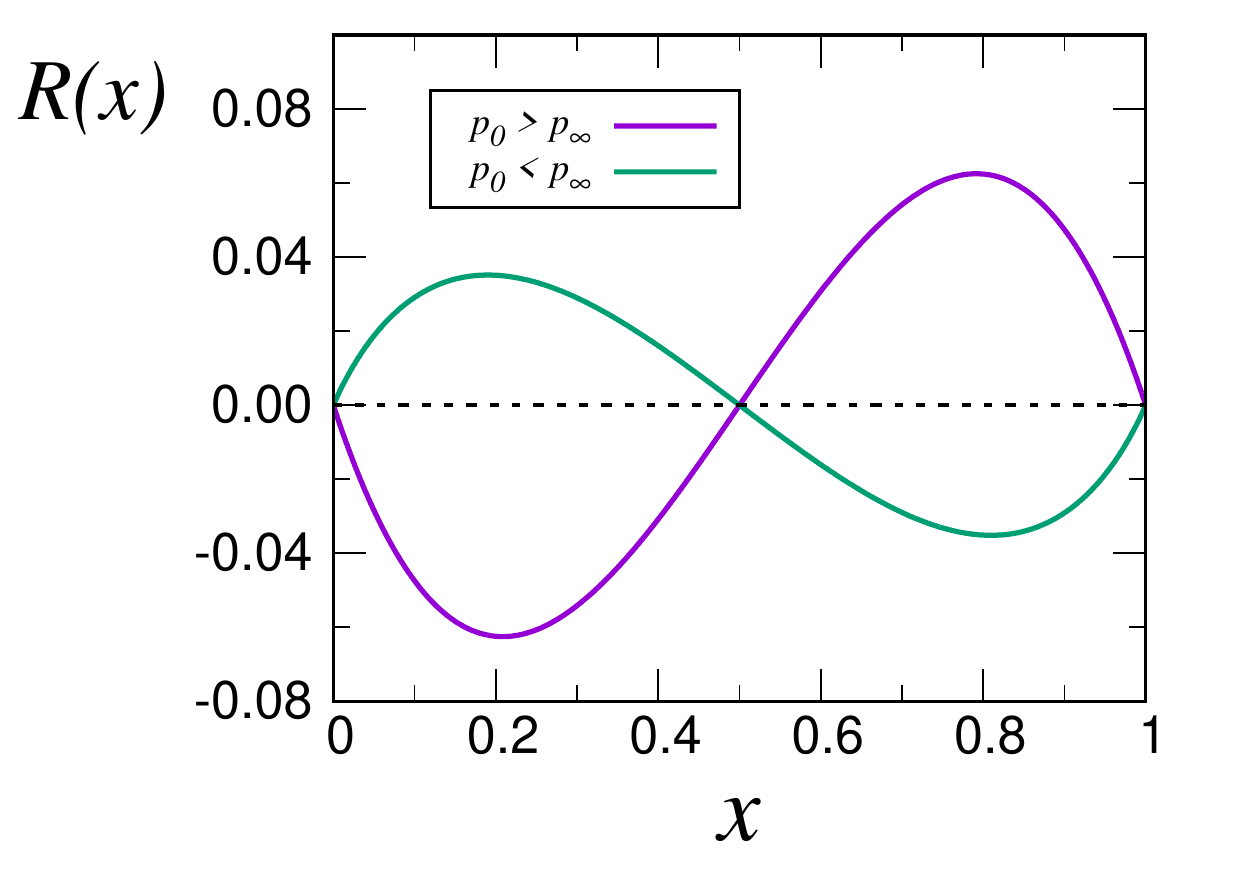}}
\subfloat[]{\label{fig:fixedb}\includegraphics[width=0.45\textwidth]{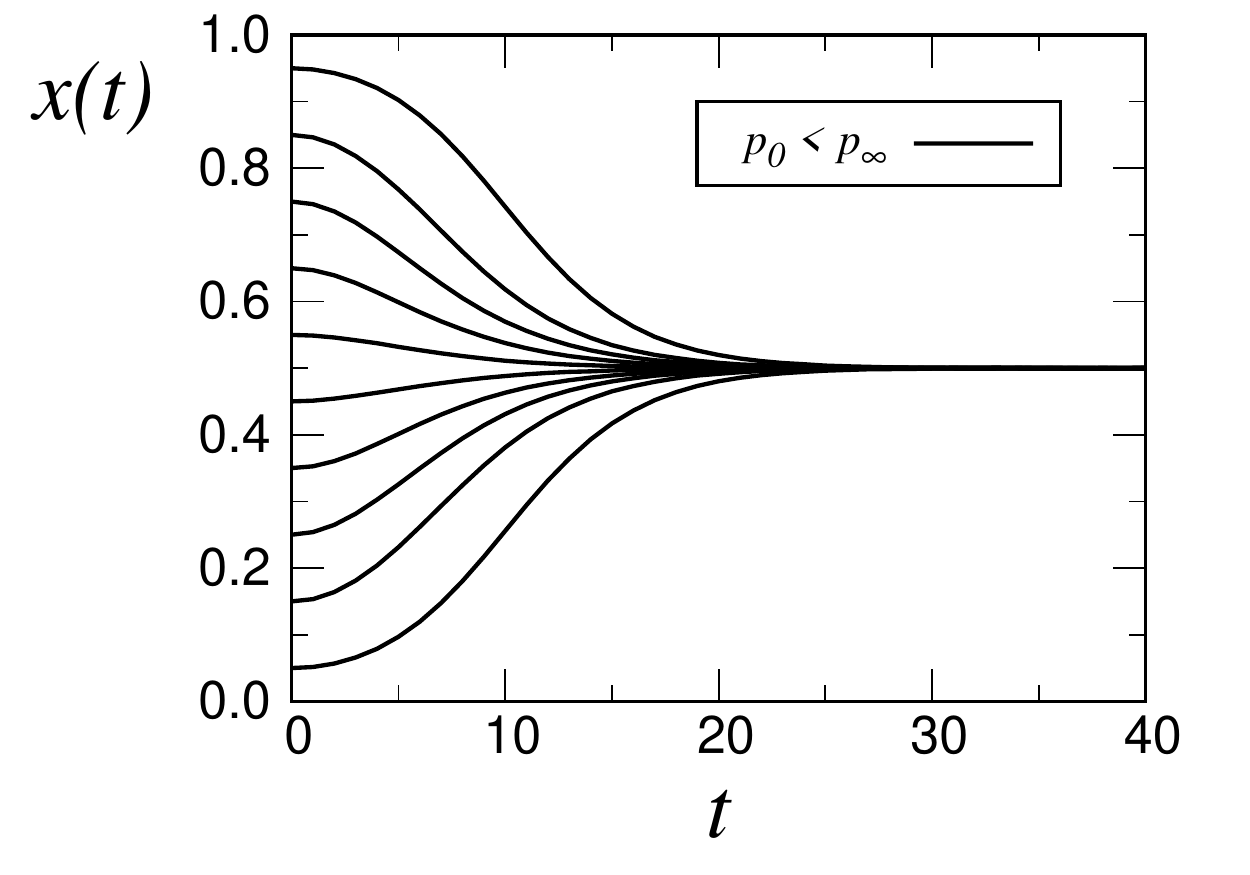}}
\caption{In the left panel we show $R(x) \equiv \tfrac{1-x}{f(1-x)}-\tfrac{x}{f(x)}$ using Eq.(\ref{eq:pv1}) for the case of aging $p_{0}=0.9$, $p_{\infty}=0.1$ and $c=1$, purple line, and for anti-aging, $p_{0}=0.1$, $p_{\infty}=0.9$ and $c=1$, green line. The right panel shows trajectories $x(t)$ coming from numerical simulations of anti-aging, $p_{0}=0.5$, $p_{\infty}=1$ and $c=1$, with different initial conditions $x(0)$ and $N=10^5$ agents, averages were performed over $10^2$ trajectories.}\label{fig:fixed} 
\end{figure}

For the activation probabilities Eq.  (\ref{eq:pv1}), one finds \footnote{Simpler expressions are obtained if $c$ is an integer number, e.g. $f(x)=\left[p_\infty(1-x)\right]^{-\frac{1-p0(1-x)}{1-p_\infty(1-x)}}$, if $c=1$.}
\begin{equation}
f(x)=1+(1-p_0(1-x)){_2F_1}\left[1,1+c\frac{1-p_0(1-x)}{1-p_\infty(1-x)};1+c;1-p_\infty(1-x)\right],
\end{equation}
where $_2F_1[\alpha,\beta; \gamma; x]$ is the Gauss hypergeometric function. This function, and hence the series Eq. (\ref{eq:funf}), is always convergent for $p_\infty(1-x)>0$. If $x=1$, we already know that the series is divergent. If $p_\infty=0$, the series diverges whenever $x\ge 1-1/(cp_0)$. Although these divergences may be problematic in the case $p_{\infty}=0$, one can still study Eq. (\ref{eq:xest}) as a limit $p_{\infty} \rightarrow 0$ or, alternatively regularize the sum Eq. (\ref{eq:funf}) with a cut-off $\sum_{i=0}^{M} (\dots)$ and study the dependence with $M$ \footnote{In some cases this limit is non-trivial and must be performed carefully. Specially when $p_{i}$ decays very fast as in the exponential example. A possibility is to consider $M(t)$ as time dependent and a phenomenological reasoning based on imposing convergence of $f(x) \sim M(t) \prod_{k}^{t} \alpha_{k}(1-x)$ with $t\rightarrow \infty$ suggests $M(t)\sim \left[ \prod_{k}^{t} (1-p_{k} x(0)) \right]^{-1}$. This is similar to imposing $\frac{d x_{i}^\pm}{d t}=0$ only for $i < t$ lower than the simulation time $t$.}. We checked that the predicted behavior of the fixed points of Eq. (\ref{eq:xest}) in all cases is correct. In Figure (\ref{fig:fixeda}) we plot $R(x) \equiv \tfrac{1-x}{f(1-x)} - \tfrac{x}{f(x)}$ in a case of aging $p_{0} >p_{\infty}$ and anti-aging $p_{0}<p_{\infty}$. We find in both cases that $R(x_{\text{st}})=0$ provides the three mentioned fixed points, $x_{\text{st}}=0,\,1/2,\,1$. We also hypothesize that the sign of $R(x)$ determines the stability or the direction of movement of $x$, $\tfrac{dx}{dt}$ , thus for aging we expect the symmetric solution $x_{\text{st}}=1/2$ to be unstable and the absorbing states $x_{\text{st}}=0, 1$ stable and the other way around for anti-aging \cite{Ozaita}. The time-dependence and ordering dynamics depending on the form of $p_{i}$ is studied in detail in the next section.

\section{Ordering dynamics}
\label{noneq}
It is clear from the rules of the process that the absorbing states $x(t)=0$ and $x(t)=1$ must be solutions of the dynamical equations in all cases, whatever the particular form of $p_i$. We now discuss these solutions. For the sake of concreteness we consider only the $x=0$ absorbing state, but an equivalent analysis could be performed at $x=1$ as well. Consider, then, that the system is at $x=0$ at time $t=0$ with all internal ages set to $0$. As there are no nodes in the $+1$ state, the herding mechanism implies that a selected node can never switch state to $+1$ and will remain in the $-1$ state, hence increasing its internal age in one unit according to the rules of the model. Therefore we have $x_{i}^{+(0)}(t)=0$, $x^{(0)}(t)=0$, and thus Eq. (\ref{eq_dyn5}) reduces to $\tfrac{dx_{0}^{-(0)}}{dt} = - x_{0}^{-(0)}$, or $x_{0}^{-(0)}(t)=e^{-t}$, where the superscript $(0)$ indicates that those values are valid at the absorbing state $x=0$. The solution of Eqs. (\ref{eq_dyn2}) with $x=0$ (implying $\alpha_j(0)=1$) and the above mentioned initial conditions can be readily found as
\begin{eqnarray}
\label{sol_eq1}
x_{i}^{-(0)}(t) = e^{-t}\frac{t^i}{i!},\quad i\ge 0,
\end{eqnarray}
which indicates that the distribution of internal ages in the population follows a Poisson distribution with $\langle i\rangle = \sigma_{i}^2 = t$. We now study the stability of this solution. 

We linearize Eqs. (\ref{eq_dyn1},\ref{eq_dyn4}) around the solution $x_i^{\pm(0)}$ at the absorbing state:
\begin{eqnarray}
\label{order1}
 \frac{dx_{i}^{+} }{dt} &=& - x_{i}^{+} + x_{i-1}^{+} (1-p_{i-1}), \quad i\ge 1,\\
\label{order2}
 \frac{dx_{0}^{+}}{dt} &=& - x_{0}^{+} + x \sum_{i=0}^{\infty} p_i x_{i}^{-(0)}=- x_{0}^{+} + x \sum_{i=0}^{\infty} p_i e^{-t}\frac{t^i}{i!}.
\end{eqnarray}
The solution of Eq. (\ref{order1}) with the initial condition $x_i^{+}(0)=0$, $i\ge1$, is
\begin{eqnarray}
\label{sol_eq2}
x_{i}^{+}(t) = \prod_{k=0}^{i-1} (1-p_k) \int_{0}^{t} \frac{(t-s)^{i-1}}{(i-1)!} e^{s-t} x_{0}^{+}(s) ds,\quad i\ge 1.
\end{eqnarray}
Imposing $x(t)=x_0^+(t)+\sum_{i=1}^\infty x_i^+(t)$, and replacing in Eq. (\ref{order2}) we obtain
\begin{eqnarray}
\label{order3}
x(t) &=& x_{0}^{+}(t) + \int_{0}^{t} G(t-s) x_{0}^{+}(s) ds, \\
\label{order4}
 \frac{dx_{0}^{+}}{dt} &=& - x_{0}^{+} + L(t) \left( x_{0}^{+} + \int_{0}^{t} G(t-s) x_{0}^{+}(s) ds \right),
\end{eqnarray}
with 
\begin{eqnarray}
\label{G_def}
G(t)&=&e^{-t} \sum_{i=0}^{\infty} \frac{t^{i}}{i!} \prod_{k=0}^{i} (1-p_k),\\
\label{L_def}
L(t)&=&e^{-t} \sum_{i=0}^{\infty} p_i \frac{t^{i}}{i!}.
\end{eqnarray}
 The integro-differential Eq. (\ref{order4}) has to be solved with the initial condition $x^+_0(0)=x(0)$. 

A simple case in which the linearized equations (\ref{order3},\ref{order4}) can be solved explicitly is $p_{i}=p$, constant. In this case we have $L(t)=p$ and $G(z)=(1-p)e^{-p z}$, which after a tedious but straightforward calculation \footnote{A possibility is to use the Laplace transform in Eqs. (\ref{order3},\ref{order4}).} leads to $x_{0}^{+}(t)=x(0)(p+(1-p)e^{-t})$ and $x(t)=x(0)$, the well known solution of the voter model in the thermodynamic limit $N\rightarrow \infty$, \footnote{Note here that $x_{0}^{+}(\infty) = p x(0)$ in apparent contradiction to the result of section \ref{sec:master}, $x_{0,\mathrm{st}}^+=px_{\mathrm{st}}(1-x_{\mathrm{st}})$. But this is a product of the linearization in Eqs. (\ref{order3},\ref{order4}).}.

As it follows from  Eqs. (\ref{order3},\ref{order4})  that $x'(0)\equiv\left.\frac{dx}{dt}\right|_{t=0}=0$, the stability of $x(t)$  requires to study the sign of $x''(0)\equiv\left.\frac{d^2x}{dt^2}\right|_{t=0}$ which after a simple calculation, turns out to be  $x''(0)=p_0(p_1-p_0)$. Therefore, at this level, we obtain that in the anti-aging case, $p_1>p_0$, the absorbing solution $x=0$ is unstable, while in the aging case, $p_1<p_0$, the absorbing solution is stable, as suggested at the end of the previous section. Since for both functional forms Eqs. (\ref{eq:pv1},\ref{eq:pv2})  in the anti-aging case there are no other solutions than $x_{\mathrm{st}}=1/2$, we predict that the final state of the dynamics is that of coexistence of states. This is indeed observed in the numerical simulations, as shown in Fig. (\ref{fig:fixedb}).  

To study in full detail the return to the absorbing state of the variable $x(t)$ we would need to solve the linearized equations (\ref{order3},\ref{order4}). Although it does not seem to be possible to obtain their full time dependence analytically \footnote{It is possible to try a power-series expansion $x(t)=\sum_{k=0}^\infty a_kt^k$ and $x_0^+(t)=\sum_{k=0}^\infty b_kt^k$ and obtain recurrence relations for the coefficients $a_k,\,b_k$, but we have not been able to sum analytically the resulting series.}, simple arguments allow us to obtain the asymptotic behaviors for large $t$. All we need for this purpose is the asymptotic expressions of $G(t)$ and $L(t)$. We split the discussion in the cases $p_\infty=0$ and $p_\infty>0$, as the difference is of crucial importance.

\subsection{Aging case, Eq. (\ref{eq:pv1}) with $p_\infty=0$.}\label{aging_power}

This case includes the one studied in \cite{Juan} where the activation probability decayed as $p_i\sim b/i$. With our explicit expression, Eq. (\ref{eq:pv1}), the calculation of Appendix \ref{app1} leads to an asymptotic behavior valid for $p_0<1$,
\begin{eqnarray}
\label{eq:asym1}
G(t)&\sim& \frac{\Gamma(c)}{\Gamma(c(1-p_0))} t^{-b},\\
L(t)&\sim&b\,t^{-1},\quad b=cp_0.
\end{eqnarray}

Assuming power-law dependences as $x_0^+(t)\sim t^{-\alpha}$, $x(t)\sim t^{-\beta}$ and that the integral in Eq. (\ref{order4}) can be approximated as $\int_{0}^{t} G(t-s) x_{0}^{+}(s) ds \sim G(t) A \sim t^{-b}$, with $A=\int_{0}^{\infty} x_{0}^{+}(s) ds$, which is justified if the time evolution of $x_0^+(t)$ is slower than that of $G(t)$, we obtain consistently $\alpha=b+1$. Finally, using this value with Eq. (\ref{order3}) we get $\beta=b$. This proves that the density of nodes in the $+$ state and, consequently, the density of active links go to zero as $x(t)\sim\rho(t)\sim t^{-b}$, explaining the numerical results of Ref. \cite{Juan}.

\begin{figure}[h!] 
\centering
\subfloat[]{\label{fig:transient:a}\includegraphics[width=0.50\textwidth]{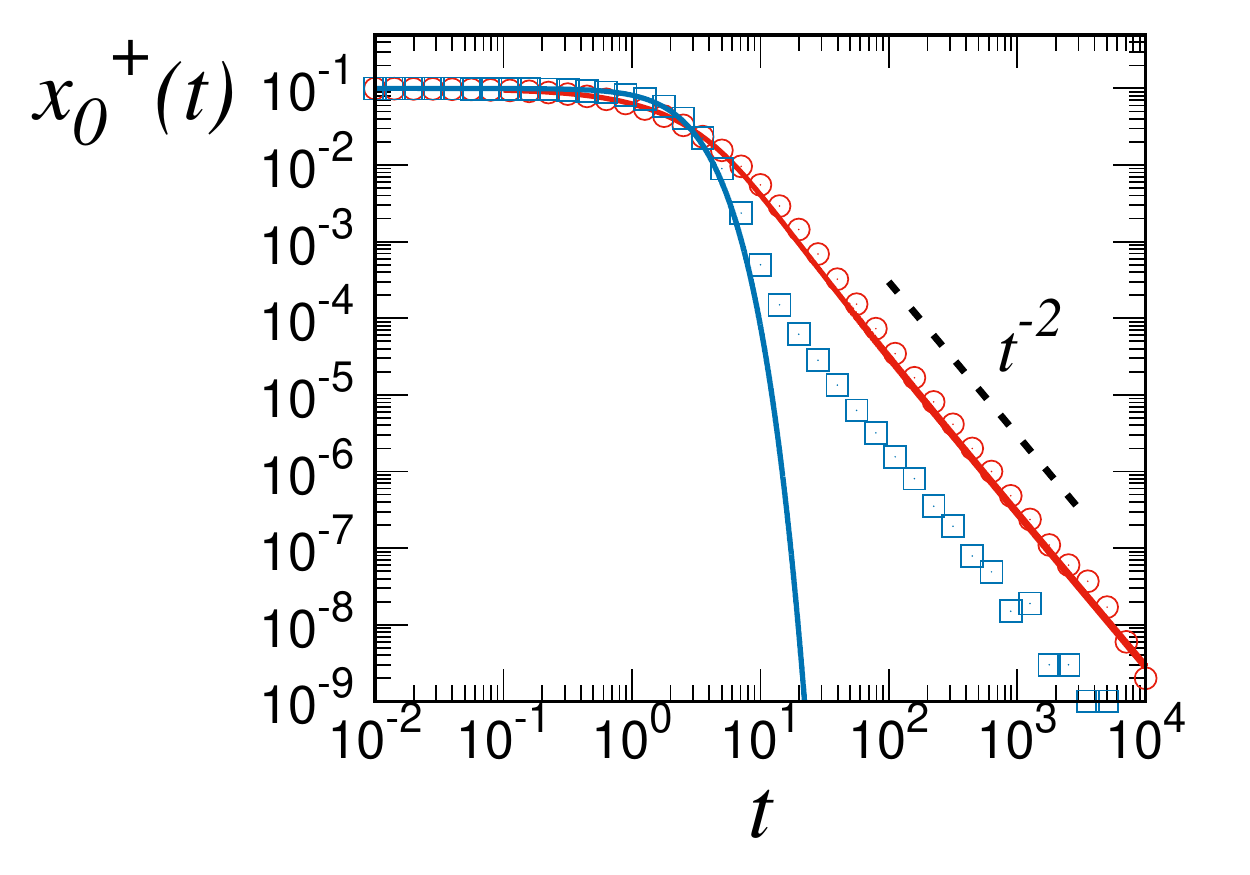}}
\subfloat[]{\label{fig:transient:b}\includegraphics[width=0.50\textwidth]{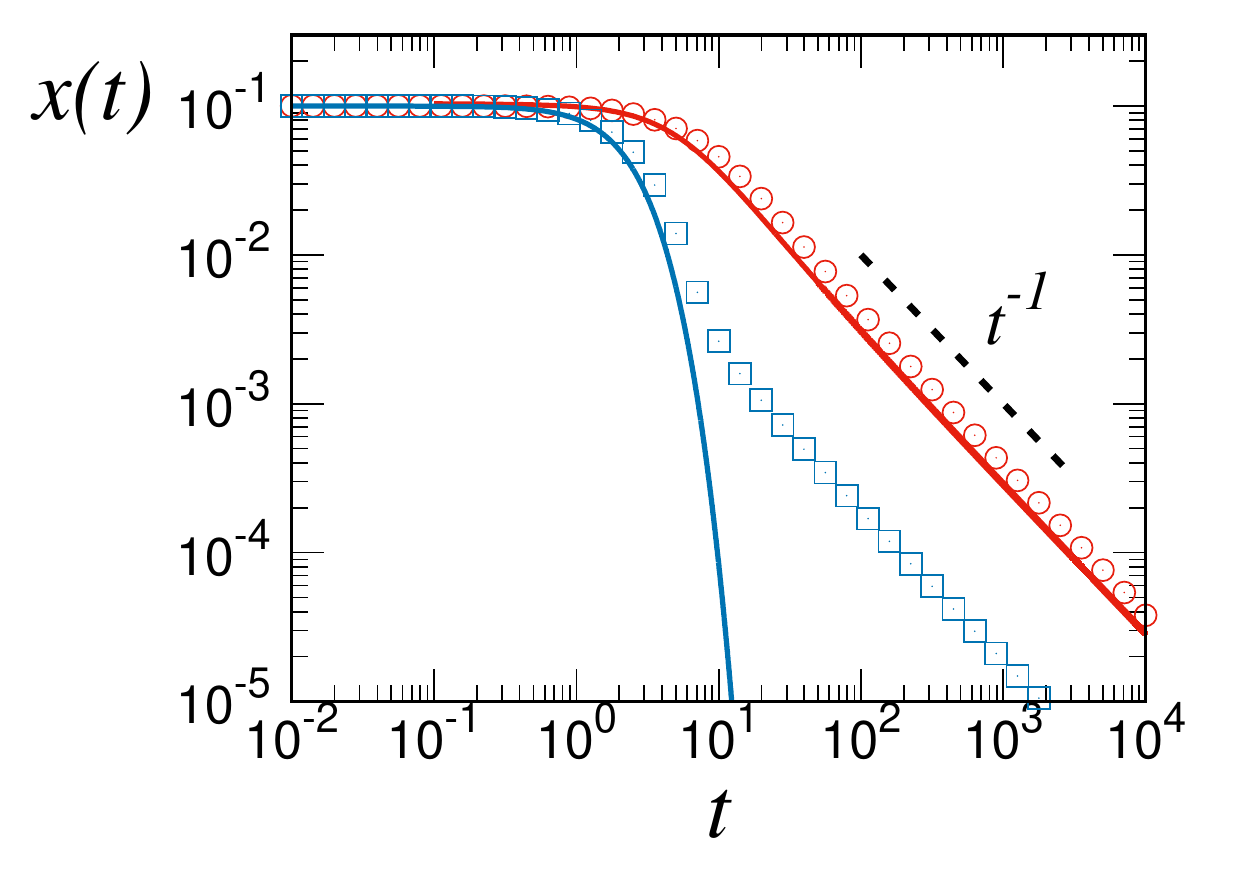}}

\subfloat[]{\label{fig:transient:c}\includegraphics[width=0.50\textwidth]{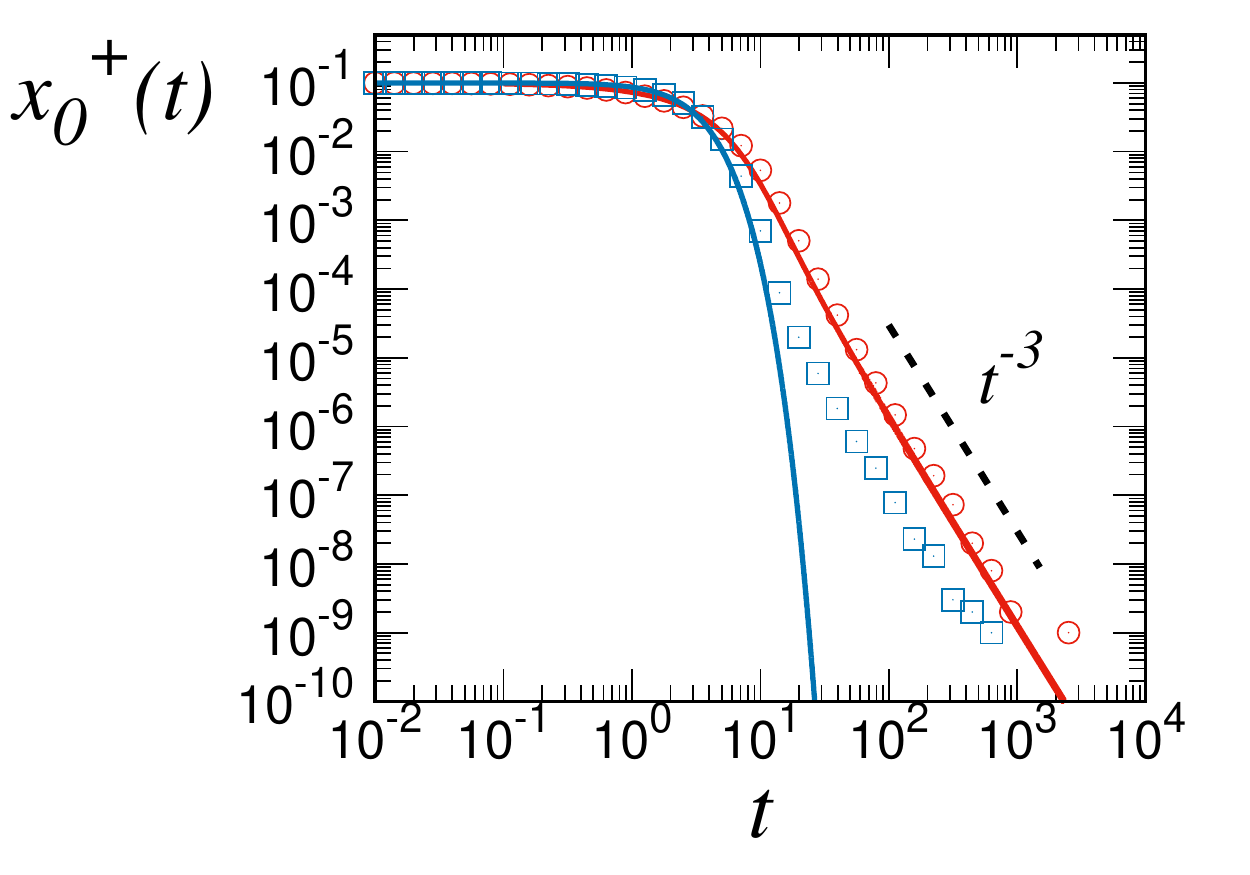}}
\subfloat[]{\label{fig:transient:d}\includegraphics[width=0.50\textwidth]{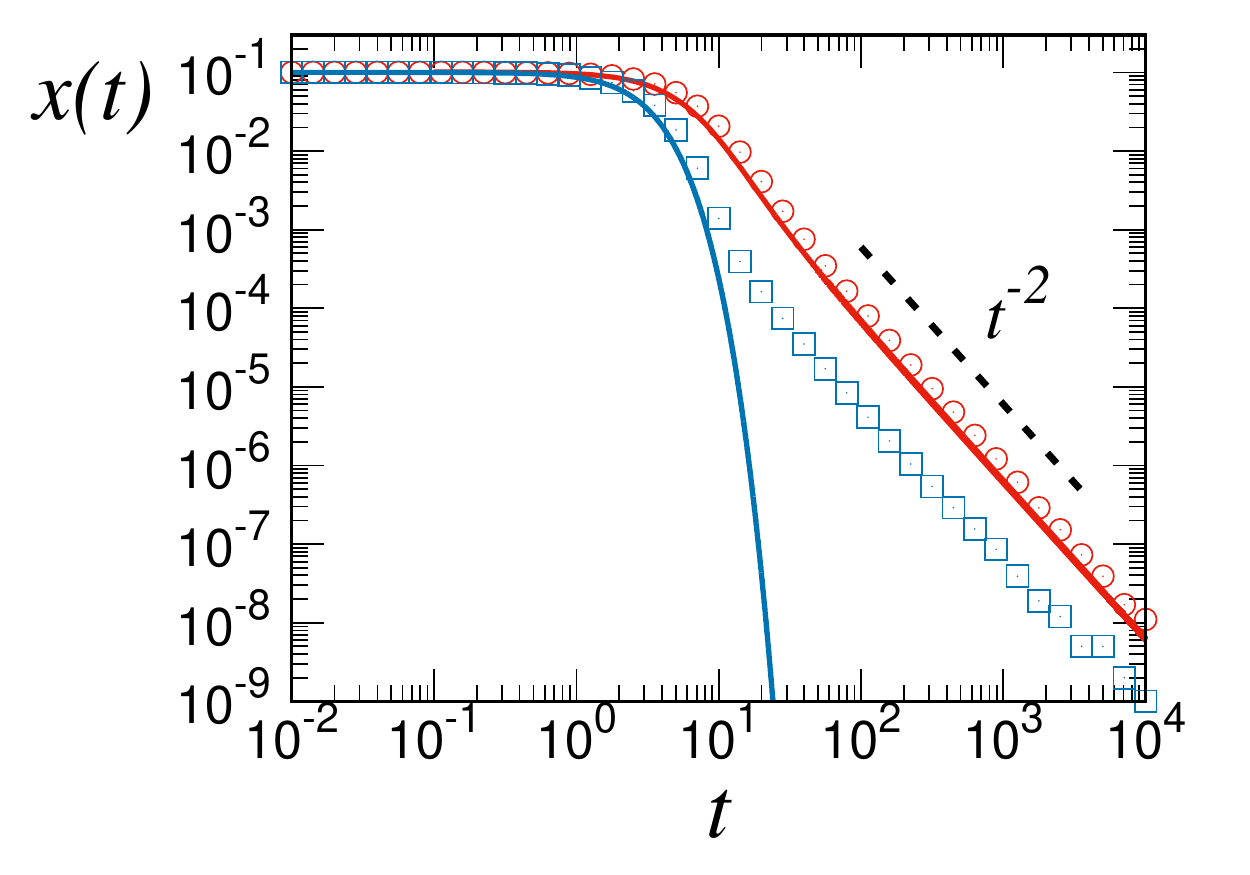}}
\caption{Time evolution of the variables $x(t), x_{0}^{+}(t)$ for $p_{i}$ given by Eq. (\ref{eq:pv1}) with different values of $p_{0}$, $c$ and $p_{\infty}=0$ and initial condition $x(0)=x_{0}^{+}(0)=0.1$. Top panels are (red circles), $p_{0}=1/2$, $c=2$, $b=1$, and (blue squares), $p_{0}=1$, $c=1$, $b=1$, and bottom panels are (red circles), $p_{0}=2/3$, $c=3$, $b=2$, and (blue squares), $p_{0}=1$, $c=2$, $b=2$. Points are results from numerical simulations with $N=10^5$ averaged over $10^4$ trajectories while lines are the numerical integration of Eqs. (\ref{order3},\ref{order4}).}\label{fig:transient} 
\end{figure}

 To check the previous predictions, we plot in Fig. (\ref{fig:transient}) the results coming from numerical simulations for $x_{0}^{+}(t)$ and $x(t)$ for different cases of $p_{0}$ and $c$. As it is apparent, the figures confirm the validity of the asymptotic expansion $x(t)\sim t^{-b}$ and $x_0^+(t)\sim t^{-b-1}$, as derived theoretically. Besides the asymptotic slope, we also plot in the figures the values of $x(t)$ and $x_0^+(t)$ obtained from the numerical solution of Eqs. (\ref{order3},\ref{order4}) using a suitable integration algorithm described in \cite{JamesThomas}.
 
This discussion assumes that $p_0<1$. If $p_0=1$, it is $G(t)=0$ and a direct integration of Eqs. (\ref{order3},\ref{order4}) leads to an initial decay $x(t)\sim e^{-t^2/(2(1+c))}$, followed by an intermediate regime $x(t)\sim t^b e^{-t}$. In the numerical simulations we observe that this is followed by a crossover to the same asymptotic power-law behavior $x(t)\sim t^{-b}$. The reason why the integro-differential Eqs. (\ref{order3},\ref{order4}) do not capture this crossover is because they are linearized. The solution of the full set of equations Eqs. (\ref{eq_dyn1}-\ref{eq_dyn3}) can be written $x(t)=x(0) h_{1}(t) + x(0)^2 h_{2}(t) + \dots$ with the different orders of the initial condition $x(0)$. Linearizing, we were able to find the first term $h_{1}(t)$. For $p_{0} \neq 1$ this first term is a power law and it prevails over the rest for all times. For $p_{0}=1$ it decays exponentially and, apparently, the second order term $h_{2}(t) \sim t^{-b}$ is a power law. Thus for sufficiently large times $t>t^{*}$ it can happen that $x(0) h_{1}(t) < x(0)^2 h_{2}(t)$.
 
\begin{figure}[h!] 
\centering
\includegraphics[width=0.6\textwidth]{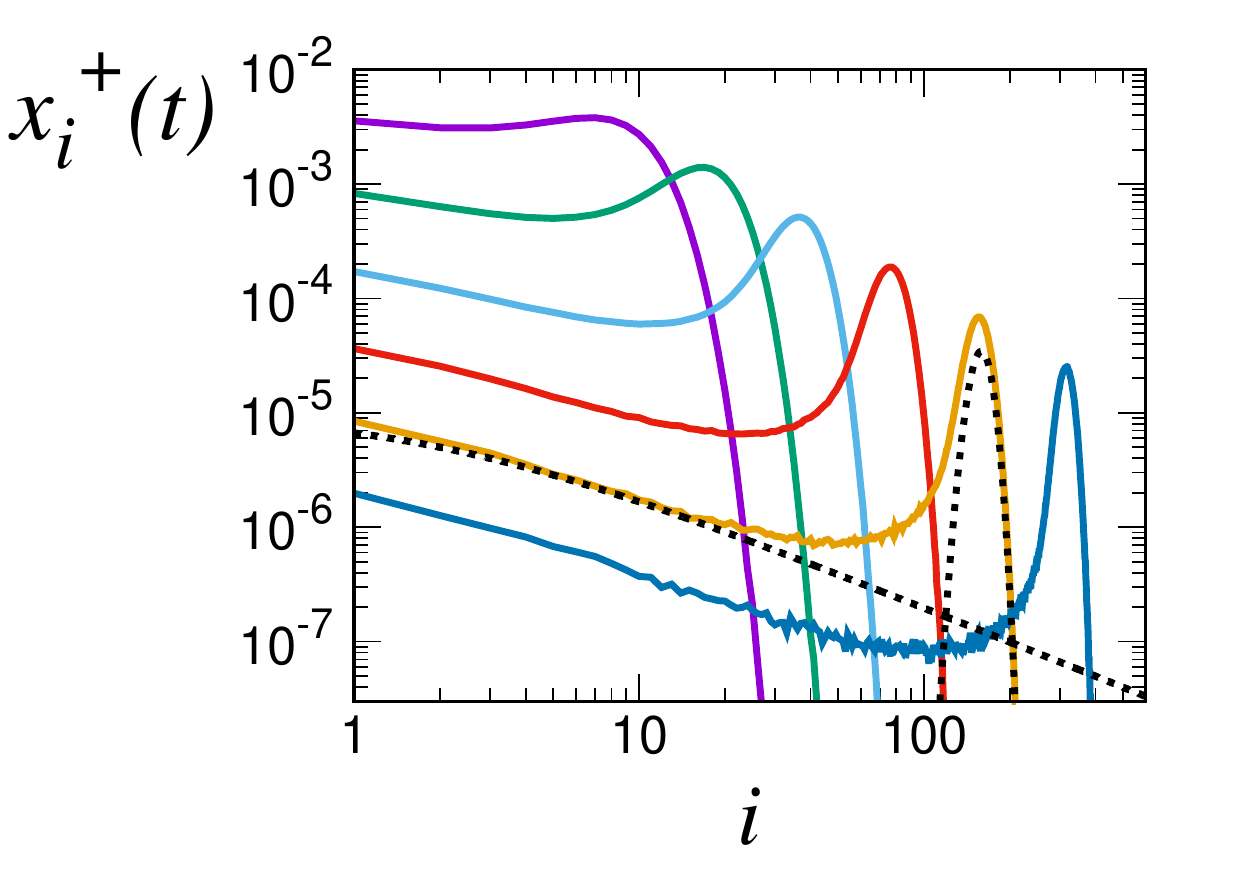}
\caption{Time evolution of the density of nodes in state $+$ with age $i$, $x_{i}^{+}(t)$ for $p_{i}$ given by Eq. (\ref{eq:pv1}) with $p_{\infty}=0$, $p_{0}=1/2$, $c=2$ with initial condition $x_{i}^{+}(0) = 0.1 \delta_{i,0}$. Colored lines are the results of numerical simulations with $N=10^5$ averaged over $10^4$ trajectories. $x_{i}^{+}(t)$ is showed as a function of $i$ for different snapshots $t=10, 20, 40, 80, 160, 320$ in log-log scale. For $t=160$ the dashed black lines are the asymptotic approximations Eqs. (\ref{order5},\ref{order6}).}\label{fig:transient2} 
\end{figure}

We now consider $x_{i}^{+}(t)$ for $i \geq 1$. The kernel $\frac{(t-s)^{i-1}}{(i-1)!} e^{s-t}$ inside the integral of Eq. (\ref{sol_eq2}) is a Poisson distribution with $\langle{i} - 1\rangle = \sigma_{i-1}^2 = t-s$, which localizes the integral of $x_{0}^{+}(s)$ in a window of position and width determined by $i$. This tells us that in order to find the qualitative shape of the age distribution $x_{i}^{+}(t)$ for $i \sim t \gg 1$, we can replace the early dynamics approximate solution $x_0^+(s)$ inside the integral at Eq. (\ref{sol_eq2}) by $x(0)e^{-\left(1-p_0\right)s}$ and $\prod_{k=0}^{i-1}(1-p_k)$ by its asymptotic expansion form $\frac{\Gamma(c)}{\Gamma(c(1-p_0))}i^{-b}$ for $i \gg 1$:
\begin{equation}
\label{order5}
x_{i\gg 1}^{+}(t) \simeq \frac{x_{0}^{+}(0)}{t^b} \frac{\Gamma(c)}{\Gamma(c(1-p_0))} \int_{0}^{t} \frac{(t-s)^{i-1}}{(i-1)!} e^{p_{0} s-t} ds.
\end{equation}
This is essentially a function localized around $i \sim t$ whose amplitude decreases as $1/t^b$. This feature of the distribution of ages is analyzed in Fig. (\ref{fig:transient2}). 

For $i \gtrsim 1$, however, the major contribution to the integral at Eq. (\ref{sol_eq2}) comes from the region $s \sim t \gg 1$ and we can use the asymptotic expression $x_{0}^{+}(s) = A s^{-(b+1)}$ as a constant inside the integral 
\begin{equation}
\label{order6}
x_{i \gtrsim 1}^{+}(t) \simeq B_it^{-(b+1)},\quad B_i\equiv A\prod_{k=0}^{i-1}(1-p_k),
\end{equation}
whose validity is checked in Fig. (\ref{fig:transient2}).

\subsection{Aging case, Eq. (\ref{eq:pv1}) with $p_\infty>0$}\label{aging_expdec}
In this case the activation probability starts at $p_0$ and decays to a non-null value $p_\infty$. This is qualitatively similar to  the case studied in \cite{Schweitzer}, although in that reference the final value $p_\infty$ was reached after a finite number of aging steps, but we do not expect this to be a significant difference. 

\begin{figure}[h!] 
\centering
\subfloat[]{\label{fig:s1a}\includegraphics[width=0.50\textwidth]{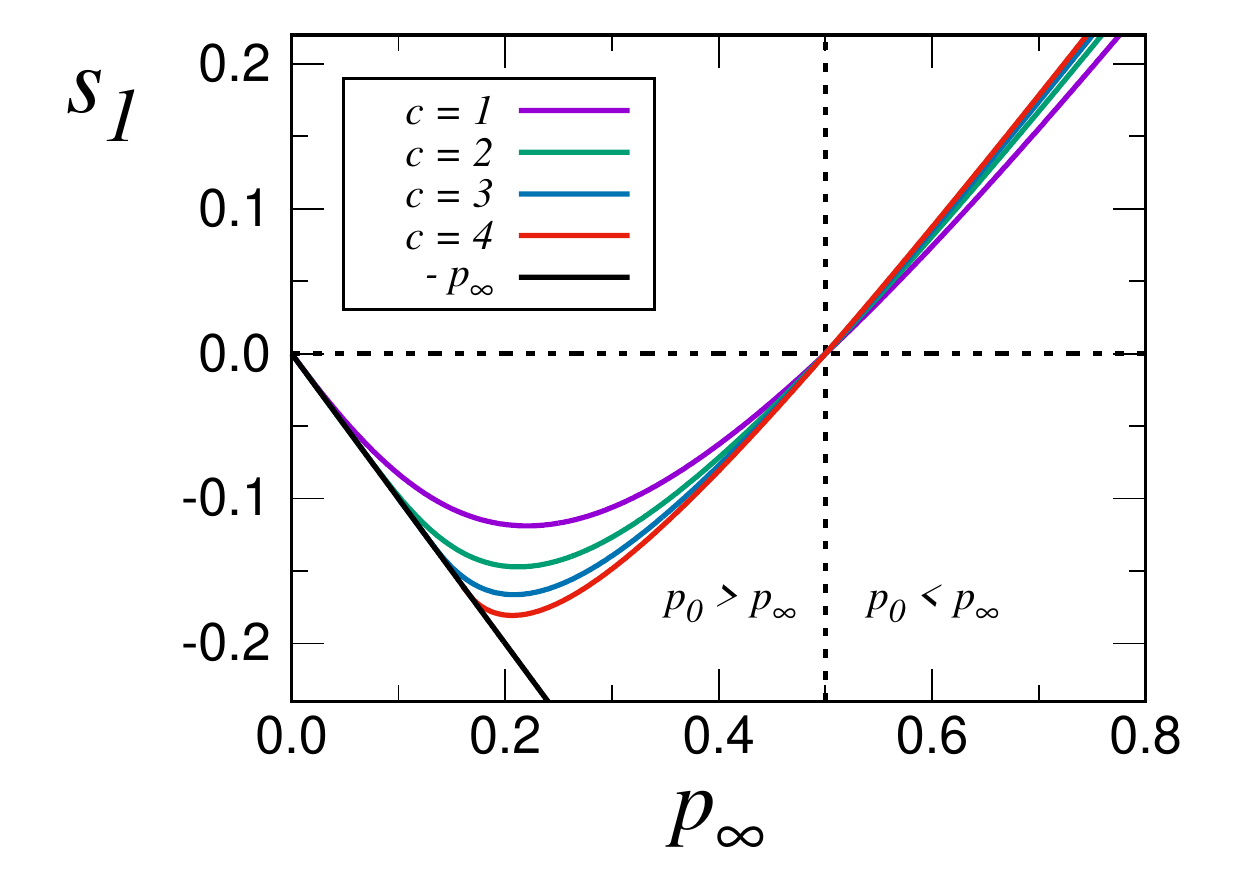}}
\subfloat[]{\label{fig:s2a}\includegraphics[width=0.50\textwidth]{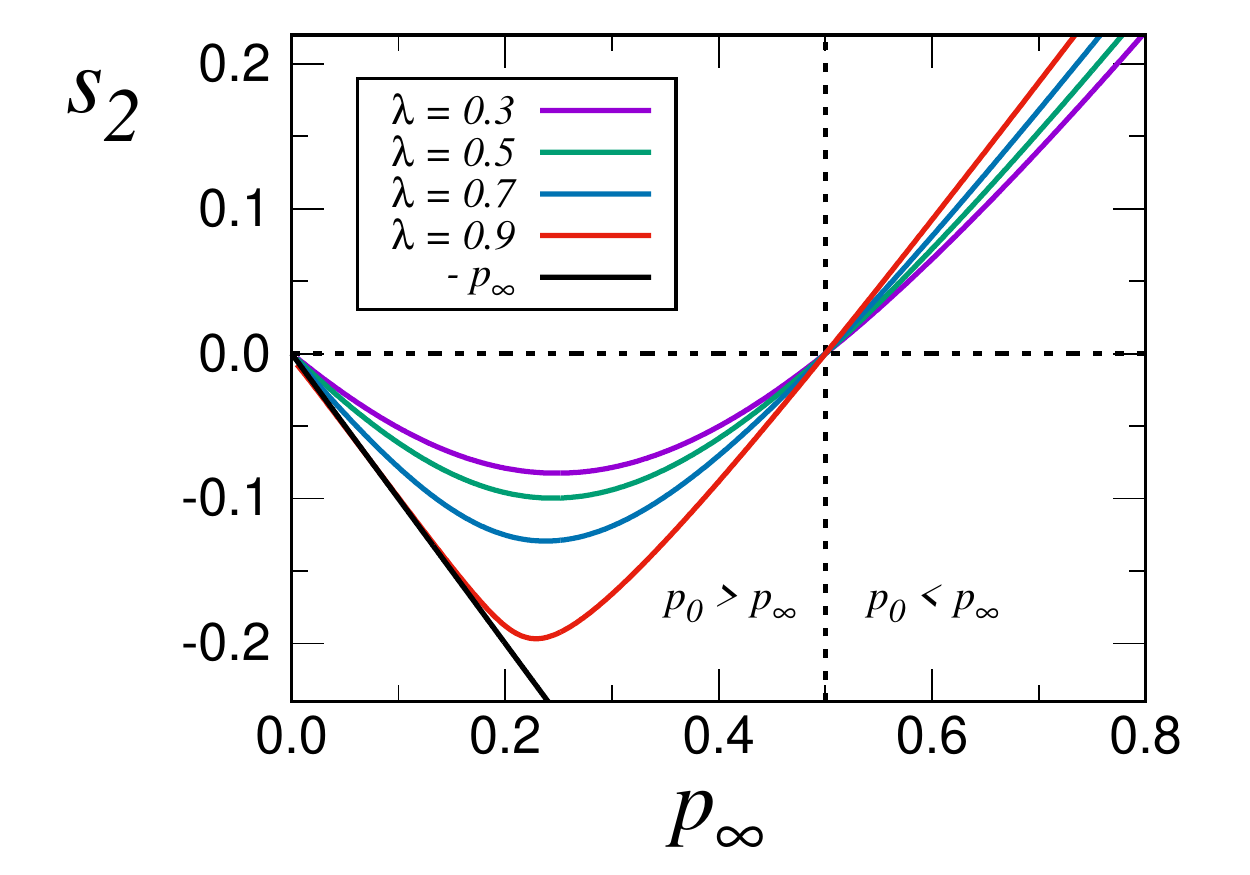}}
\caption{Roots of the equation $\hat{F}(s)=s+1-p_\infty(1+\hat G(s))=0$ as a function $p_{\infty}$ with fixed $p_{0}=0.5$. The right panel is the case $p_{i}$ given by Eq. (\ref{eq:pv1}) for different values of $c=1,2,3,4$ and in the left panel it is Eq. (\ref{eq:pv2}) for different values of $\lambda=0.3,0.5,0.7,0.9$. The explicit forms of the Laplace transform of $G$ are given in Appendix \ref{app1}.}\label{fig:s1} 
\end{figure}

\begin{figure}[h!] 
\centering
\subfloat[]{\label{fig:transient3:a}\includegraphics[width=0.50\textwidth]{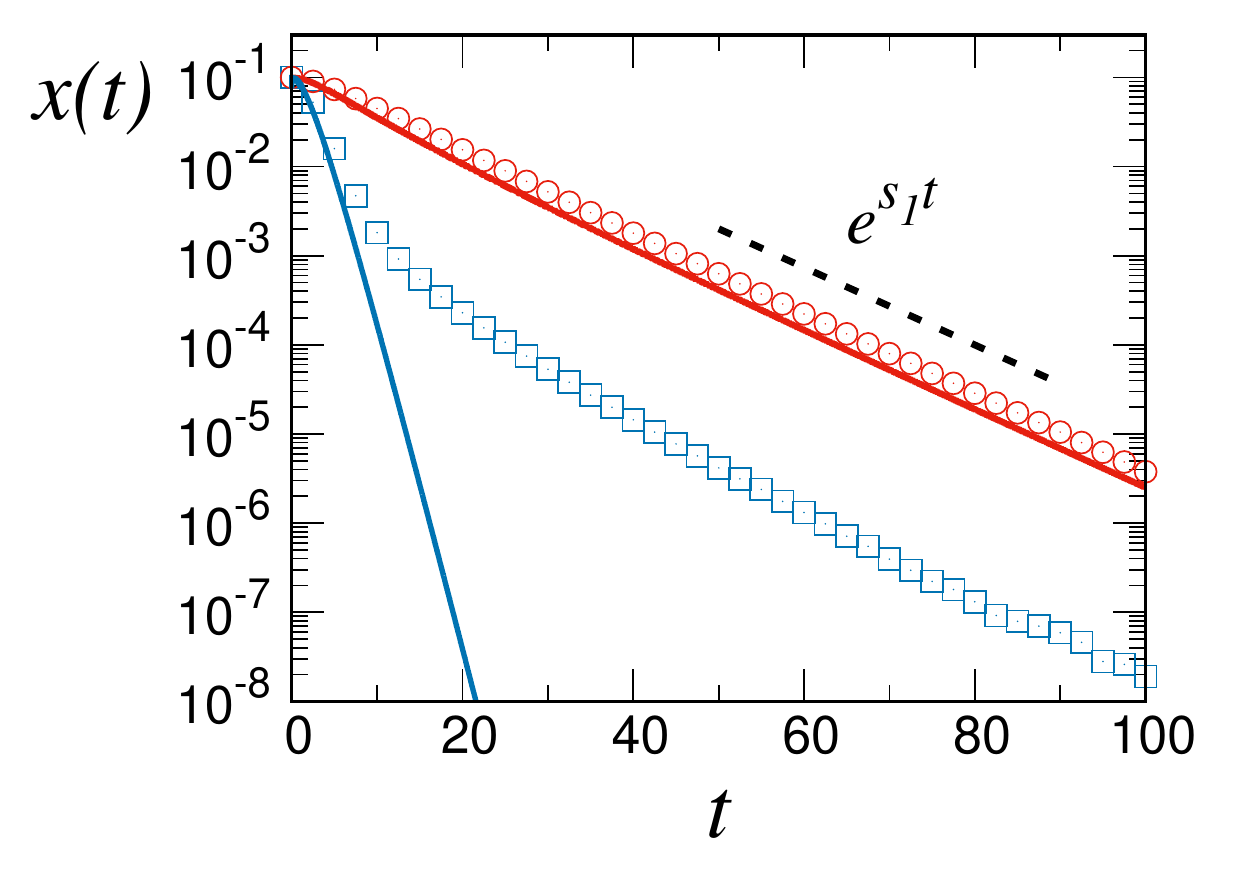}}
\subfloat[]{\label{fig:transient3:b}\includegraphics[width=0.50\textwidth]{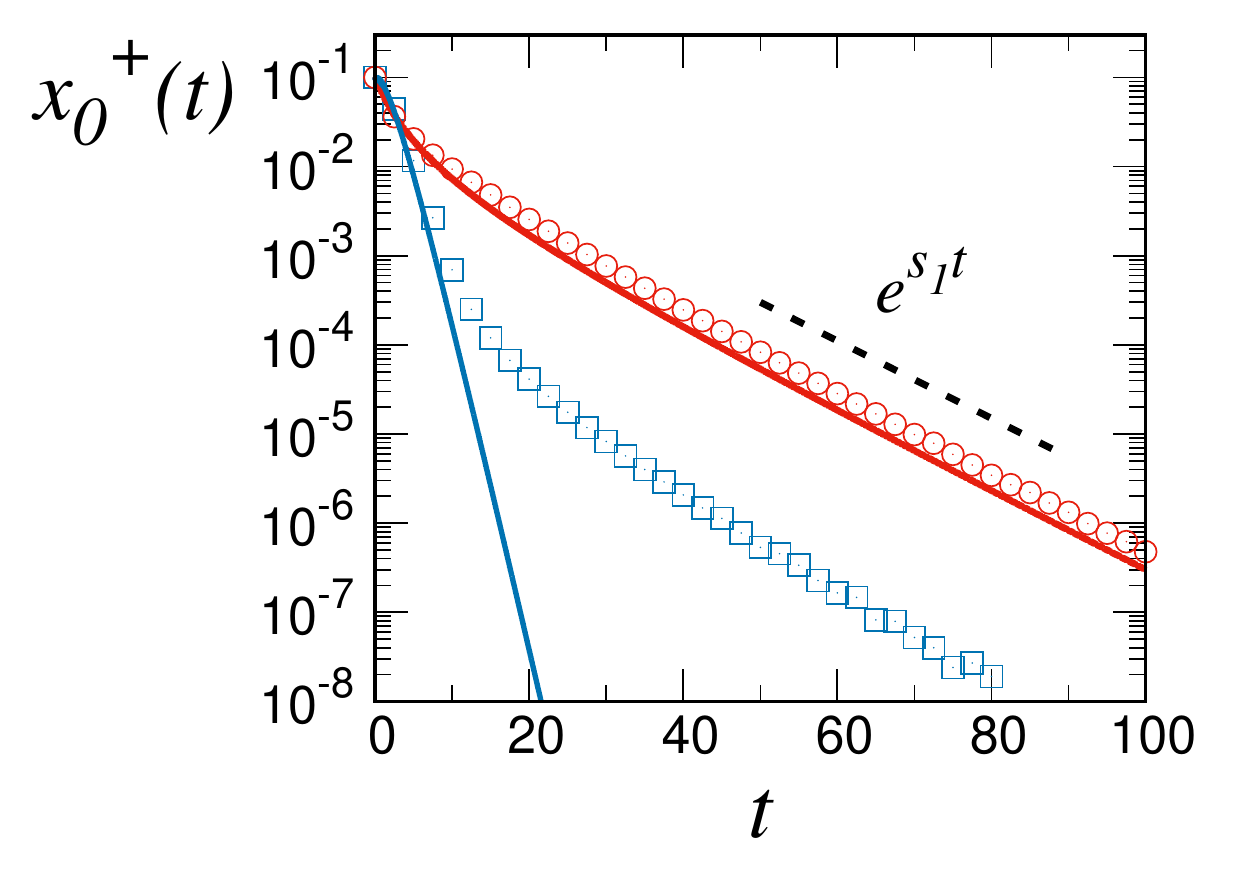}}
\caption{Time evolution of the variables $x(t), x_{0}^{+}(t)$ for $p_{i}$ given by Eq. (\ref{eq:pv1}) with different values of $p_{0}$, $c$ and $p_{\infty}=0.1$ and initial condition $x(0)=x_{0}^{+}(0)=0.1$. Red circles are $p_{0}=1/2$, $c=1$, and blue squares, $p_{0}=1$, $c=1$. Points are results from numerical simulations with $N=10^5$ averaged over $10^4$ trajectories, while solid lines are the numerical integration of Eqs. (\ref{order3},\ref{order4}).}\label{fig:transient3} 
\end{figure}

As shown in Appendix \ref{app1}, $L(t)$ tends to a non-null asymptotic value $L(t\to\infty) = p_\infty$ while $G(t)$ decays exponentially:
\begin{eqnarray}
\label{eq:asym2}
G(t) &\sim& e^{-p_{\infty} t} t^{-b},\quad b = c \frac{p_0-p_{\infty}}{1-p_{\infty}}\\
L(t) &\sim&  p_\infty.
\end{eqnarray}
If we replace $L(t)$ by this constant value $p_\infty$ in Eq. (\ref{order4}) we can use the Laplace transform $\hat {x}_0^+(s)$ of $x_0^+(t)$ to find the explicit solution:
\begin{equation}
\label{Laplace_x0}
\hat {x}_0^+(s)=\frac{x(0)}{s+1-p_\infty(1+\hat G(s))}\equiv\frac{x(0)}{\hat F(s)},
\end{equation}
and from Eq. (\ref{order3}) we get the Laplace transform $\hat x(s)$ of $x(t)$:
\begin{equation}
\label{Laplace_x}
\hat x(s)=\frac{1+\hat G(s)}{\hat F(s)}x(0).
\end{equation}
Note that, due to the asymptotic behavior of $G(t)$, Eq. (\ref{eq:asym2}), the Laplace transform $\hat{G}(s)$ does not converge for $s < - p_{\infty}$. It turns out that $\hat F(s)$ has a zero located at $s=s_1$. Hence, inverting the Laplace transform we find the asymptotic behavior $x_0^+(t)\sim x(t)\sim e^{s_1t}$. The exact value of $s_1$ has to be found solving numerically $\hat F(s_1)=0$ and it is found to depend non-trivially on the parameters of the model, see Figure (\ref{fig:s1a}). As it is shown in this figure, we find $s_{1}<0$ for $p_{0}>p_{\infty}$ (aging) and $s_{1}>0$ for $p_{0}<p_{\infty}$ (anti-aging), thus the asymptotic behavior is in accordance to the linear stability analysis performed before. There are two cases where $s_{1}=0$: (i) for $p_{0}=p_{\infty}$ which corresponds to the non-aging voter model with $x(t)=x(0)$, and (ii) for $p_{\infty}=0$ where the time dependence changes from being exponential to a power law $x(t) \sim t^{-\beta}$ as explained in the previous section. The case that concern us in this section is $0 < p_{\infty} < p_{0}$, where we have $s_{1} \sim -p_{\infty}$ for small $p_{\infty}$ and as $p_{\infty}$ approaches $p_{0}$, $s_{1}$ increases until $s_{1}=0$. We have checked the proposed asymptotic behavior in the particular case $p_{\infty}=0.1$, $p_{0}=0.5$, $c=1$ in Figure \ref{fig:transient3}. The theory predicts an exponential decay with $s_{1}=-0.0982$ which matches perfectly compared to numerical simulation and also compared to the numerical solution of Eqs. (\ref{order3},\ref{order4}). In the same figure we also show the special problematic case $p_{0}=1$, with the same other parameters. In this case it is $G(t)=0$ thus according to Eqs. (\ref{order3},\ref{order4}) the solution is initially an exponential function as $x(t) \approx e^{-\frac{1-p_{\infty}}{2(1-c)}t^2}$ followed by the regime $x(t) \approx e^{-(1-p_{\infty})t} t^{p_{0} c}$. This behavior is correct in the early states of the dynamics but after it there is a crossover to a slower exponential decay $x(t) \sim e^{s_1^{*} t}$. We expect this effect to be also a result of linearization, where in $x(t)=x(0) h_{1}(t) + x(0)^2 h_{2}(t) + \dots$, $h_{2}(t)$ is slower than $h_{1}(t)$ which is the result that we obtained. Thus for a large enough time $t > t^{*}$ it can happen that $x(0) h_{1}(t) < x(0)^2 h_{2}(t)$.

\subsection{Aging case, Eq. (\ref{eq:pv2}) with $p_\infty=0$}
\label{case5.3}
\begin{figure}[h!] 
\centering
\subfloat[]{\label{fig:transient4:a}\includegraphics[width=0.50\textwidth]{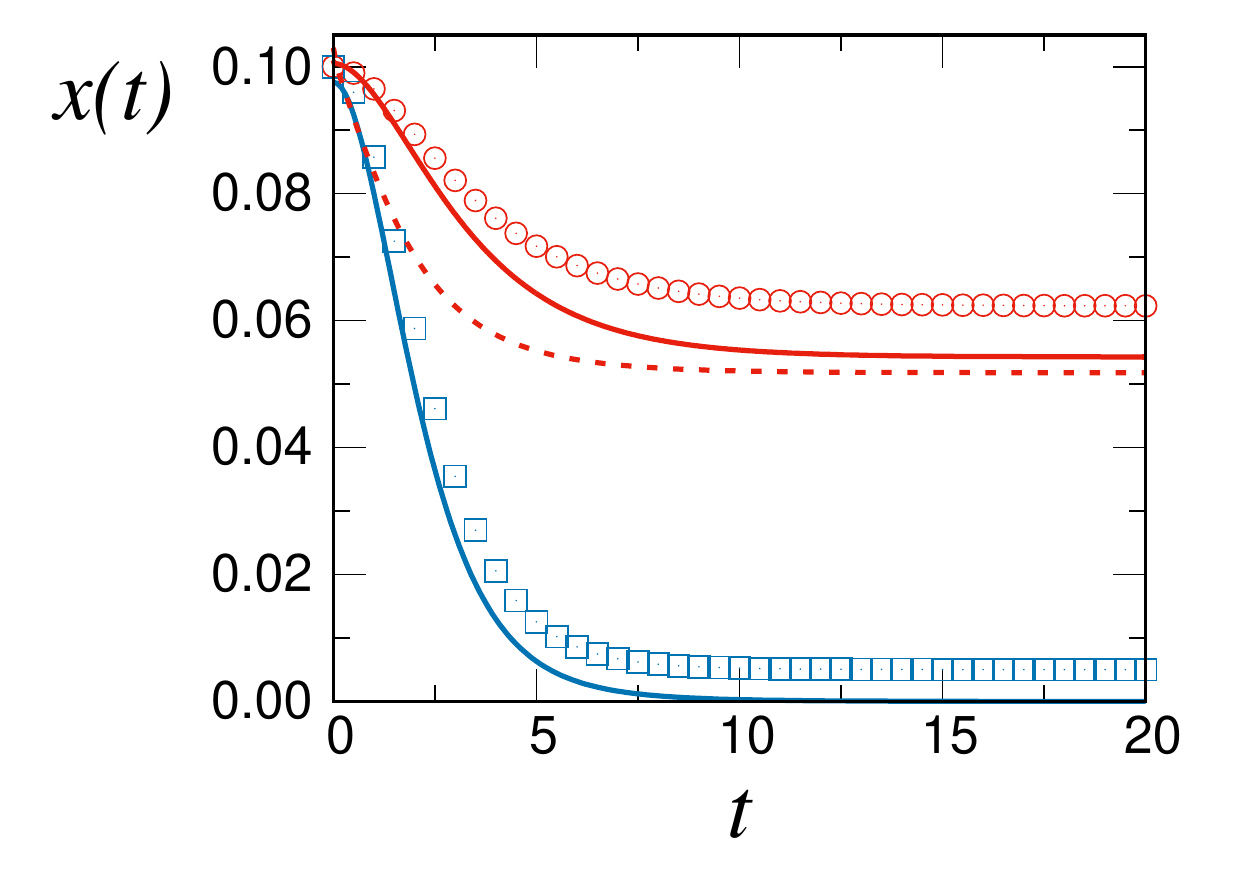}}
\subfloat[]{\label{fig:transient4:b}\includegraphics[width=0.50\textwidth]{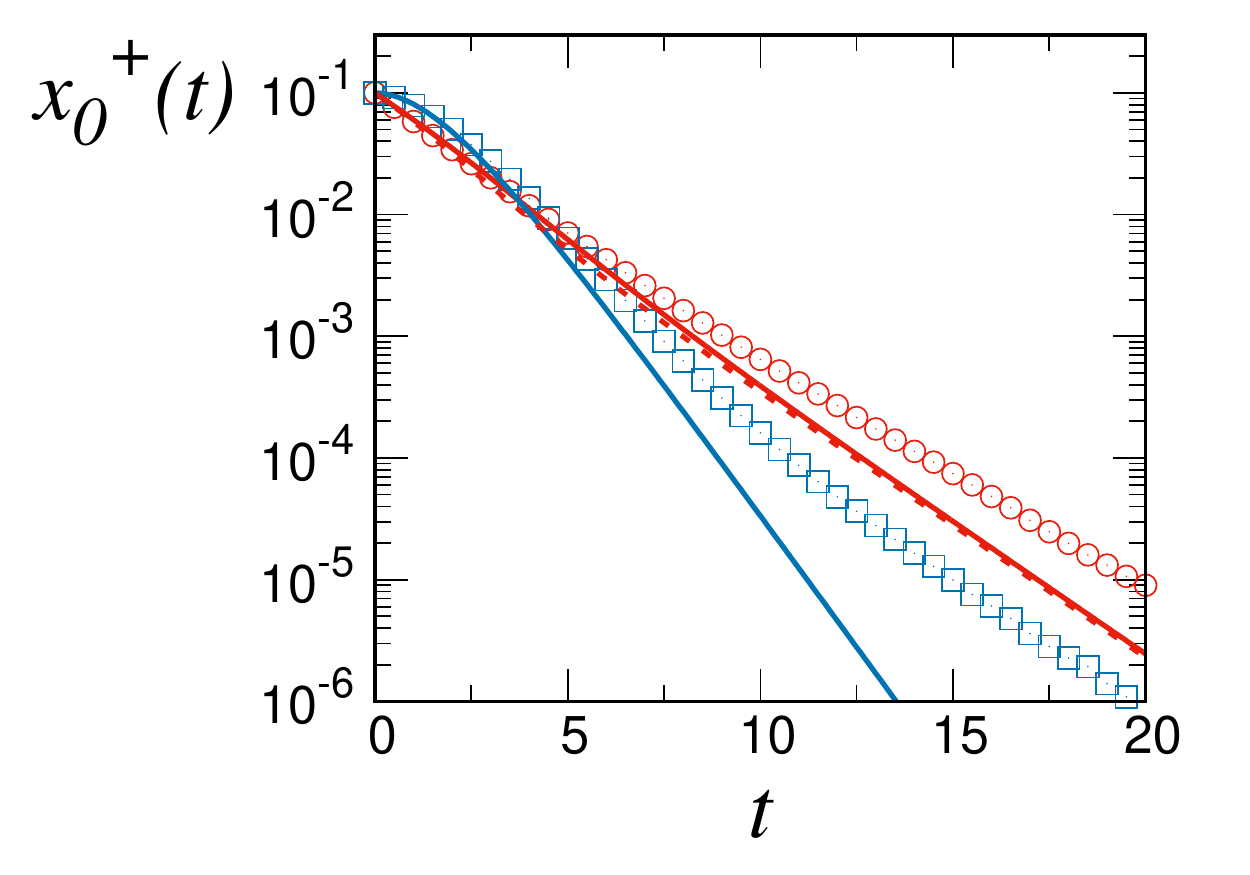}}
\caption{Time evolution of the variables $x(t), x_{0}^{+}(t)$ for $p_{i}$ given by Eq. (\ref{eq:pv2}) with different values of $p_{0}$, $\lambda$ and $p_{\infty}=0$ and initial condition $x(0)=x_{0}^{+}(0)=0.1$. Red circles are $p_{0}=1/2$, $\lambda=1/2$, and blue squares, $p_{0}=1$, $\lambda=1/2$. Points are results from numerical simulations with $N=10^5$ averaged over $10^4$ trajectories, while lines are the numerical integration of Eqs. (\ref{order3},\ref{order4}). The dashed lines are the approximate solution of Appendix \ref{appb}.}\label{fig:transient4} 
\end{figure}

\begin{figure}[h!] 
\centering
\includegraphics[width=0.50\textwidth]{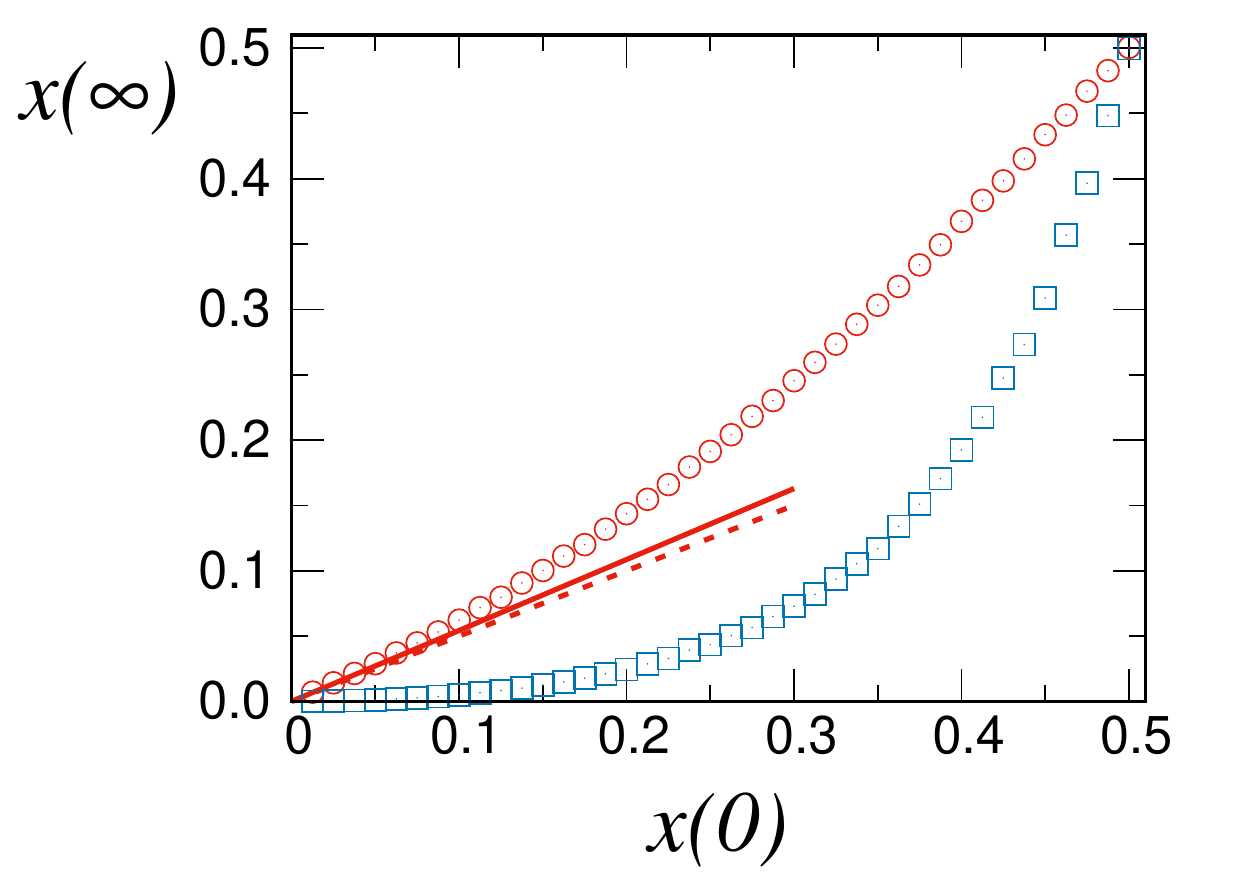}
\caption{Stationary value $x(\infty)$ as a function of the initial condition $x(0)$ for $p_{i}$ given by Eq. (\ref{eq:pv2}) with $\lambda=1/2$, $p_{\infty}=0$ and $N=10^5$. Red circles are $p_{0}=1/2$ and blue squares $p_{0}=1$, while the red line is the theoretical prediction of the linearized Eqs. (\ref{order3},\ref{order4}), and the dashed line is the approximation of Appendix \ref{appb}, i.e. $x(\infty) \approx c_2\lambda/p_0 x(0)$.}\label{fig:fixed2} 
\end{figure}

From expression Eq. (\ref{eq:pv2}) in Appendix \ref{app1} we find the asymptotic behavior for large $t$
\begin{eqnarray}
\label{eq:asym3}
G(t)&\sim&  \left(p_0;\lambda\right)_\infty,\\
\label{eq:asym3b}
L(t)&=&p_0e^{-(1-\lambda)t},
\end{eqnarray}
where $(a;b)_{i} \equiv \prod_{k=0}^{i-1} (1-a b^{k})$ is the q-Pochhammer symbol \cite{qpochhammer}.
Note the remarkable property that $G(t \rightarrow \infty) \rightarrow \text{constant}>0$ which is something that we did not observe in the other cases. As explained in Appendix \ref{appb}, we can solve Eqs. (\ref{order3},\ref{order4}) in this case replacing Eqs. (\ref{eq:asym3},\ref{eq:asym3b}). Approximately, we have the exponential decay $x_{0}^{+}(t) \approx x_{0}(0) c_2e^{-(1-\lambda)t}$ and $x(t)$ decays exponentially until it reaches a constant value $x(\infty) \approx x(0)c_2\lambda/p_0 $  that depends on the initial condition $x(0)$. Here $c_2$ is a constant that depends on $p_0$ and $\lambda$ in an intricate way, see Appendix \ref{appb}. In summary, the system does not order in this case. We can deduce that this will be the case as long as $G(\infty) = \prod_{k=0}^{\infty} (1-p_{k}) \neq 0$ or, equivalently, if $\sum_{k=0}^{\infty}p_{k}$ converges. Obviously for $p_{\infty} \neq 0$ this is never the case, while for $p_{\infty} = 0$ if $p_{k}$ decays faster than $p_{k} \sim 1/k$ the system does no order, but it does order if $p_{k}$ decays slower than $p_{k} \sim 1/k$. The case $p_{k} \sim 1/k$ is somehow a critical value, where the system orders very slowly as a power law as explored in section \ref{aging_power}. We checked this prediction against the results of  numerical simulations in Figure \ref{fig:transient4} for $p_{0}=0.5$, $\lambda=1/2$. For $p_{0}=1$ there is again the failure of the linearization,  where the theory predicts that the system orders exponentially, while the numerical simulation shows the same behavior initially until a crossover is reached to a constant value $x(\infty) \neq 0$, which is smaller than in the case $p_{0} \neq 1$. According to this argument $x(t)=x(0) h_{1}(t) + x(0)^2 h_{2}(t) + \dots$, $h_{1}(\infty)=0$ and $h_{2}(\infty) \neq 0$. The fixed point value is expected to scale as $x(\infty) \propto x(0)^2$ as the linear theory predicts $x(\infty) = 0$. This and the linear prediction $x(\infty) \approx c_2\lambda/p_0 x(0)$ is checked with numerical simulation in Figure \ref{fig:fixed2}.

\subsection{Aging case, Eq. (\ref{eq:pv2}) with $p_\infty>0$}
\begin{figure}[h!] 
\centering
\subfloat[]{\label{fig:transien5t:a}\includegraphics[width=0.50\textwidth]{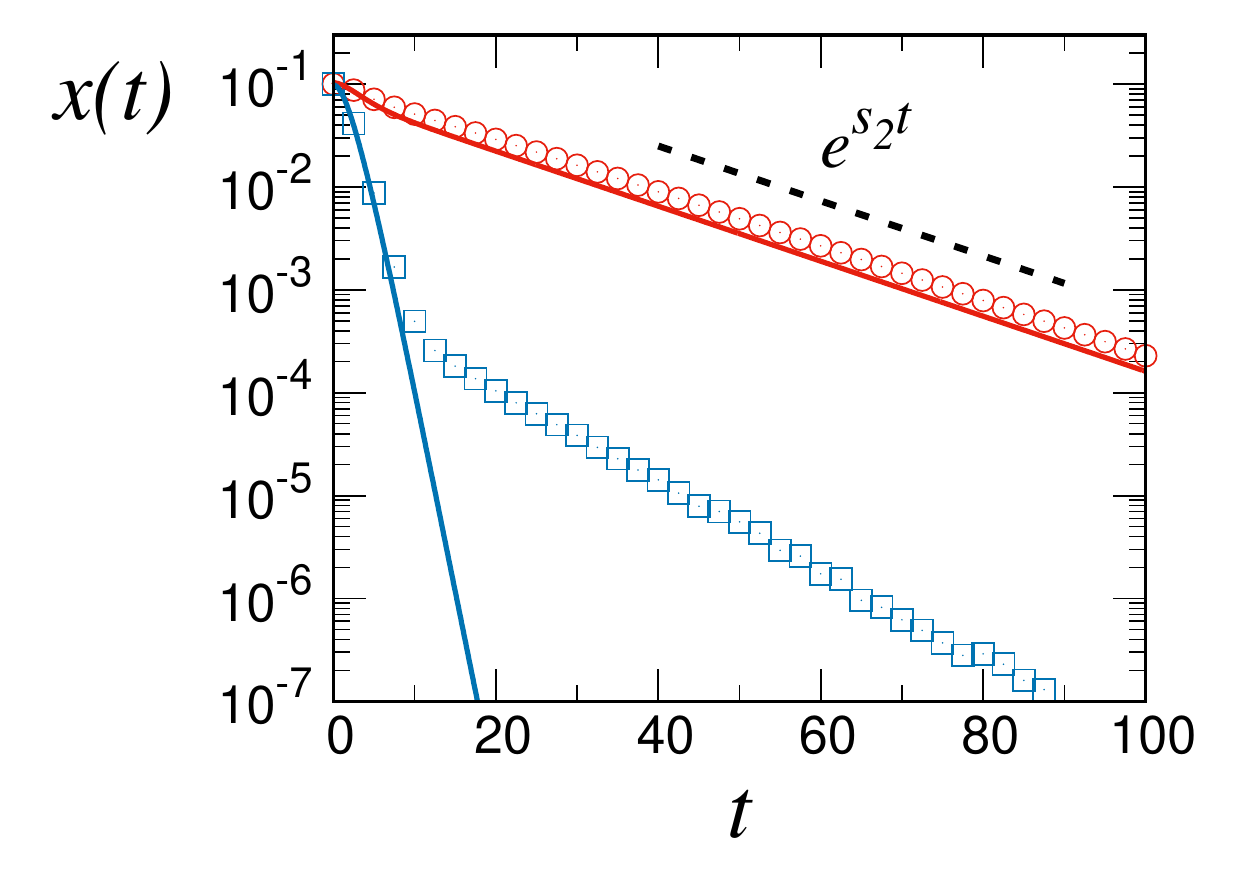}}
\subfloat[]{\label{fig:transient5:b}\includegraphics[width=0.50\textwidth]{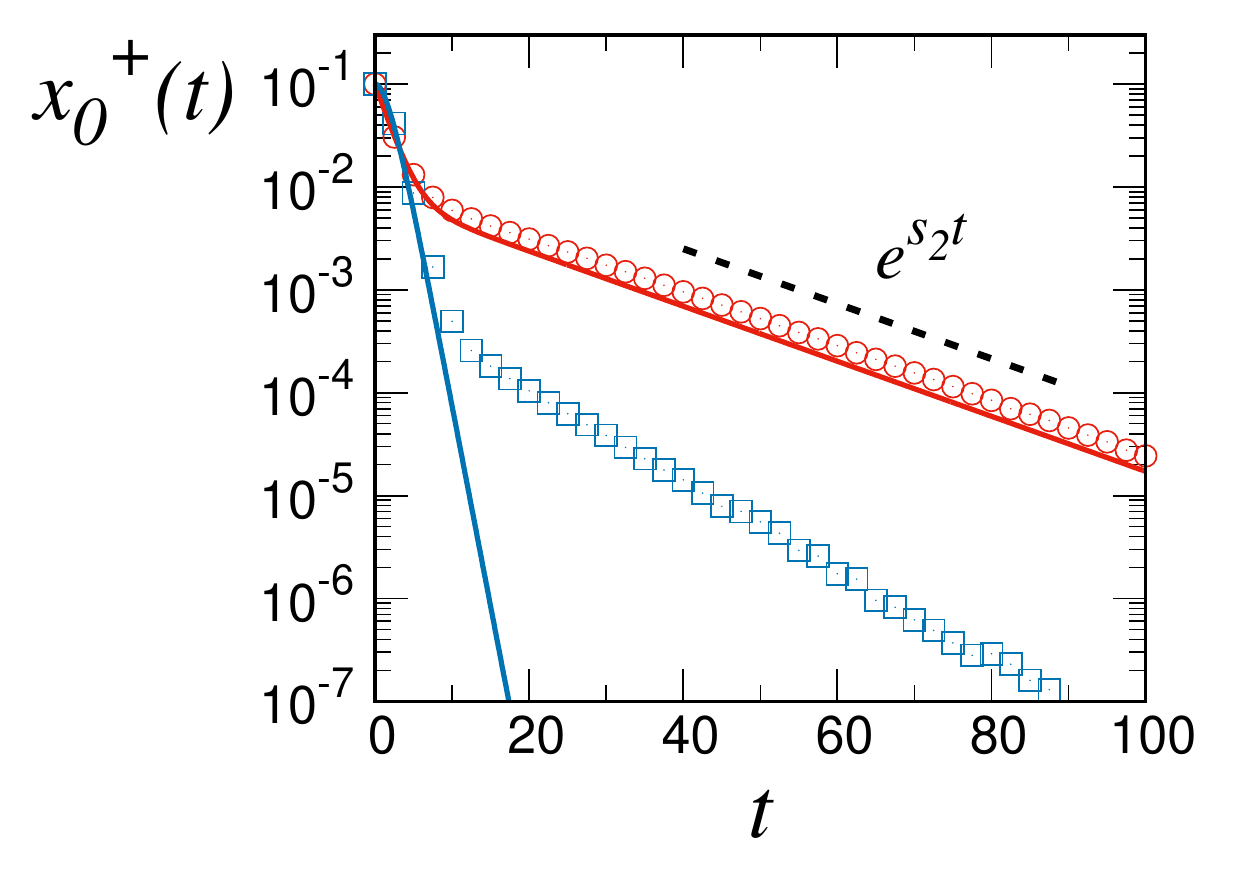}}
\caption{Time evolution of the variables $x(t), x_{0}^{+}(t)$ for $p_{i}$ given by Eq. (\ref{eq:pv2}) with different values of $p_{0}$, $\lambda$ and $p_{\infty}=0.1$, and initial condition $x(0)=x_{0}^{+}(0)=0.1$. Red circles are $p_{0}=1/2$, $\lambda=1/2$, and blue squares, $p_{0}=1$, $\lambda=1/2$. Points are results from numerical simulations with $N=10^5$ averaged over $10^4$ trajectories, while lines are the numerical integration of Eqs. (\ref{order3},\ref{order4}).}\label{fig:transient5} 
\end{figure}

In this case, from expression Eq. (\ref{eq:pv2}) in Appendix \ref{app1} we find the asymptotic expressions
\begin{eqnarray}
\label{eq:asym4}
G(t)&\sim&  (1-p_\infty)\left(\frac{p_0-p_\infty}{1-p_\infty};\lambda\right)_{\infty}e^{-p_\infty t},\\
L(t)&\sim&p_\infty.
\end{eqnarray}
Thus, this case is similar to the one studied in section \ref{aging_expdec} where $G(t)$ decays exponentially and $L(t)$ goes to a constant. Following the procedure of that section we use the Laplace transform in Eq. (\ref{order3},\ref{order4}) after replacing $L(t)$ by its limit $p_{\infty}$, see Appendix \ref{app1} for an explicit expression for these functions. In Figure (\ref{fig:s2a}) we explored the root $s_2$ of the denominator $\hat{F}(s)$ of Eq. (\ref{Laplace_x0}) as a function of $p_{\infty}$. We find the same phenomenology, with $s_{2}>0$ for $p_{0} < p_{\infty}$ and $s_{2}<0$ for $p_{0} > p_{\infty}$, $s_{2}=0$ for $p_{0} = p_{\infty}$ corresponding to the non-aging voter model. For $p_{\infty}=0$ we also obtain $s_{2}=0$, in accordance to the result of the previous section where $x(t)$ goes to a constant. In Figure \ref{fig:transient5} we compare the results of numerical simulations of $x(t)$ with the numerical integration of Eqs. (\ref{order3},\ref{order4}) and the asymptotic $x(t) \sim x_{0}^{+}(t) \sim e^{s_{2} t}$, for $p_{\infty}=0.1$, $\lambda=0.5$, $p_{0}=0.5$ and $p_{0}=1$. The theory predicts $s_{2}=-0.0616\dots$ which is in good agreement to what it is observed in the simulation. The case $p_{0}=1$ is again singular as the linear approximation fails to reproduce the long term dynamics. In the linear theory we have $G(t)=0$ which leads to $h_{1}(t)\sim e^{-(1-p_{\infty})t+\frac{p_{0}-p_{\infty}}{1-\lambda}(1-e^{-(1-\lambda)t})}$. This exponential decay happens to be faster than the second order contribution, which is also exponential $h_{2}(t) \sim e^{s_{2}^{*} t}$, and thus for long enough time the second order prevails.

\section{Summary and conclusions}
\label{summary_conclusions}

In this work we have studied memory effects (aging and anti-aging) in an opinion dynamics model, by means of analytical and numerical approaches. More specifically, we have focused on a mean-field (all-to-all) description of the voter model, where the influence of the aging of the individuals has been modelled through the activation probability, $p_{i}$, or the copying probability of the individuals as a function of their internal age $i$. Overall, we have shown that the functional form of $p_{i}$ has a crucial impact on how and whether the system reaches consensus or not. For the sake of concreteness, we have employed in our analysis monotonic functions $p_{i}$ with three parameters: $p_{0}$ (or $p_i$ for zero age), $p_{\infty}$ (or $p_i$ for $\infty$ age), and a third one, $c$ or $\lambda$, tuning the transition $p_{0} \rightarrow p_{\infty}$. 

When the activation probability is constant $p_0=p_\infty$, we recover the voter model. In this case, the system keeps its initial magnetization, provided the number of agents is large enough. If the activation probability is not constant and tends to a nonzero value for large ages, $p_0\ne p_\infty >0$, the aging and anti-aging cases produce opposite effects in the dynamics. On the one hand, for the case of aging, with $p_{0} > p_{\infty}>0$, the system orders with an exponential decay of the global opinion towards one of the consensus states $x \rightarrow 0, 1$. On the other hand, for the anti-aging case, with $0<p_{0} < p_{\infty}$, the system reaches coexistence with $x \rightarrow 1/2$. 

As explained in Ref. \cite{Schweitzer}, the results for $p_\infty>0$ can be understood heuristically: the aging mechanism amplifies any small asymmetry in the initial conditions. Take an initial condition close to consensus, say $x=0$, and all the agents with zero age. Then, the systems evolves in a first, very short stage as if no aging was present since all agents have the same herding rate. After the first transitions, most of the agents become older in the state $-1$, while a small fraction are younger with states $\pm 1$. From a dynamical point of view, the only difference between the latter situation and the voter model is in the presence of the small fraction of the young agents. In the case of aging, since the rate of changing state is bigger for the young agents, the most probable event is that a young agent copies the state of an old one. Altogether, the magnetization decreases. On the contrary, for the case of anti-aging with the young agents having smaller rates, the most probable event is that of an old copying a young. Now, the magnetization increases. 

For the case of aging with zero $p_{\infty}$,  $p_{0} > p_{\infty}=0$, the system may or may not order, depending on how ``fast'' is the decay of $p_i$ to zero as the age increases. From the theoretical analysis of the integro-differential equations close to the consensus state, together with the results of numerical simulations, we infer the following criterion: if $p_{i}$ decays faster than $1/i$ then the system gets trapped in a frozen state given by the initial conditions, while the system orders in other cases. The decay $p_{i} \sim 1/i$ can be then regarded as a critical functional form where we proved that the system orders with a power-law time dependence. 

The results of $p_\infty =0$ can also be understood using qualitative arguments. The evolution of the magnetization can be explained as for the case $p_\infty >0$, the only difference now being the time needed for the system to reach its final value. For $p_i$ decaying slow enough, the system reaches consensus but on a time much larger than that for $p_\infty>0$, since it takes more time to convince all agents, specially the older ones. For the other case of $p_i$ decaying faster, almost all agents become zealots \cite{Nagi} on a short time scale, and the system gets frozen on an active state, $x\ne 0,1$, in general. 

\appendix
\section{Asymptotic expansions}
\label{app1}
\subsection{}
If we use the expression Eq. (\ref{eq:pv1}) for the activation probabilities, we obtain the explicit expressions for the functions Eqs. (\ref{G_def},\ref{L_def})
\begin{eqnarray}
\label{eq:Gz}
G(t)&=&(1-p_0)e^{-t}M\left(1+c\frac{1-p_0}{1-p_\infty},1+c,(1-p_\infty)t\right),\\
\label{eq:Lz}
L(t)&=&e^{-t}\left[p_0M\left(c,c+1,t\right)+t\frac{p_\infty}{1+c}M\left(1+c,2+c,t\right)\right],
\end{eqnarray}
where $M(\alpha,\beta,x)$ is Kummer's $_1F_1(\alpha;\beta;x)$ confluent hypergeometric function. Using the expansion for large $z$: $M(\alpha,\beta,z)\sim \frac{\Gamma(\beta)}{\Gamma(\alpha)}e^z z^{\alpha-\beta}$ we obtain in the asymptotic limit $t\to\infty$ the expansions:
\begin{eqnarray}
\label{eq:Gza}
G(t)&\sim&Ae^{-p_\infty t}t^{-b}, \\
\label{eq:Lza}
L(t)&\sim&\frac{p_0 c}{t}+p_\infty,\\
\quad A&=&(1-p_{\infty})\frac{\Gamma(c)}{\Gamma\left(c\frac{1-p_0}{1-p_\infty}\right)},\\
 b&=&c\frac{p_0-p_\infty}{1-p_\infty}.
\end{eqnarray}
The expressions for $G(t)$ assume $p_\infty\ne 1$. For $p_\infty=1$ we find
\begin{eqnarray}
\label{eq:Gz1}
G(t)&=&\Gamma(c)(c(1-p_0))^{1-c/2}e^{-t}t^{-c/2}I_c\left(2\sqrt{c(1-p_0)t}\right)\\
&\sim&\frac{\Gamma(c)}{\sqrt{2\pi\sqrt{2}}}[c(1-p_0)]^{(3-2c)/4}t^{-(c+2)/4}e^{-t}.
\end{eqnarray}

It is also possible to obtain the Laplace transform of $G(t)$ as:
\begin{equation}
\hat G(s)=\frac{1-p_0}{1+s}{_2F_1}\left[1,1+c\frac{1-p_0}{1-p_\infty};1+c;\frac{1-p_\infty}{1+s}\right].
\end{equation}

\subsection{}
In the case of activation probabilities given by Eq. (\ref{eq:pv2}) it is possible to obtain $L(t)$ analytically:
\begin{equation}
\label{eq:Lz2}
L(t)=(p_0-p_\infty)e^{-(1-\lambda)t}+p_\infty,
\end{equation}
although we have not been able to find a closed expression for $G(t)$, 
\begin{equation}
\label{Gssum_e}
G(t)=e^{-t}\sum_{i=0}^\infty\frac{(1-p_\infty)^{i+1}\left(\frac{p_0-p_\infty}{1-p_\infty};\lambda\right)_{i+1}}{i!}t^i,
\end{equation}
with $(q;a)_n$ the q-Pochhammer symbol. It is possible, however, to obtain the asymptotic expansion for $t\to\infty$ as
\begin{equation}
\label{Gssum}
G(t)\sim (1-p_\infty)\left(\frac{p_0-p_\infty}{1-p_\infty};\lambda\right)_{\infty}e^{-p_\infty t}.
\end{equation}
In the particular case $p_\infty=0$ it yields $G(t)\sim \left(p_0;\lambda\right)_{\infty}$.

Finally, we mention the Laplace transform of $G(t)$ as:
\begin{equation}
\label{hatGs}
\hat G(s)=\sum_{i=0}^\infty\left(\frac{1-p_\infty}{1+s}\right)^{i+1}\left(\frac{p_0-p_\infty}{1-p_\infty};\lambda\right)_{i+1}.
\end{equation}
For the numerical calculation of $G(t)$ and $\hat G(s)$ in this case we use Eqs.(\ref{Gssum_e},\ref{hatGs}) using a sufficiently large number of terms in the sums.

\section{Solution of Eqs. (\ref{order3},\ref{order4}) using Eqs. (\ref{eq:asym3},\ref{eq:asym3b})}
\label{appb}
If $G(t)\equiv G_\infty$ is a constant, we can differentiate Eq. (\ref{order3}) and obtain a system of differential equations:
\begin{eqnarray}
\label{order3b}
\frac{dx}{dt} &=& \frac{dx_{0}^{+}}{dt} + G_\infty x_{0}^{+}(t), \\
\label{order4b}
 \frac{dx_{0}^{+}}{dt} &=& - x_{0}^{+} + L(t) x(t).
\end{eqnarray}
From the last equation we get $x(t)=\left(\frac{dx_{0}^{+}}{dt}+x_{0}^{+}\right)/L(t)$ which, replaced in Eq.(\ref{order3b}) leads to a closed second order linear differential equation for $x_{0}^{+}(t)$. The change of variables $z\equiv -\frac{p_0}{1-\lambda}e^{-(1-\lambda)t}$ makes the coefficients of the equation to be simple polynomials and a standard technique based of Frobenius power-series expansion leads to the general solution $x(t)=\left[c_1x^{(1)}(t)+c_2x^{(2)}(t)\right]x(0)$, $x_0^+(t)=\left[c_1x_0^{(1)}(t)+c_2x_0^{(2)}(t)\right]x(0)$, where $c_1$ and $c_2$ are constants imposed by satisfying the initial condition $x(0)=x_0^+(0)=1$, and
\begin{eqnarray}
x_0^{(1)}(t)&=&e^{-t}M\left(\frac{1-G_\infty}{1-\lambda},\frac{1}{1-\lambda},-\frac{p_0}{1-\lambda}e^{-(1-\lambda)t}\right),\\
x_0^{(2)}(t)&=&e^{-(1-\lambda)t}M\left(1-\frac{G_\infty}{1-\lambda},\frac{1-2\lambda}{1-\lambda},-\frac{p_0}{1-\lambda}e^{-(1-\lambda)t}\right),\\
x^{(1)}(t)&=&(1-G_\infty)e^{-t}M\left(1+\frac{1-G_\infty}{1-\lambda},\frac{2-\lambda}{1-\lambda},-\frac{p_0}{1-\lambda}e^{-(1-\lambda)t}\right),\\
x^{(2)}(t)&=&\frac{\lambda}{p_0}M\left(1-\frac{G_\infty}{1-\lambda},\frac{1-2\lambda}{1-\lambda},-\frac{p_0}{1-\lambda}e^{-(1-\lambda)t}\right)+\\
&&\frac{1-\lambda-G_\infty}{1-2\lambda}e^{-(1-\lambda)t}M\left(2-\frac{G_\infty}{1-\lambda},\frac{2-3\lambda}{1-\lambda},-\frac{p_0}{1-\lambda}e^{-(1-\lambda)t}\right).
\end{eqnarray}
In the limit $t\to\infty$, the asymptotic behavior is
\begin{eqnarray}
x_0^{(1)}(t)&\rightarrow&e^{-t},\\
x_0^{(2)}(t)&\rightarrow&e^{-(1-\lambda)t},\\
x^{(1)}(t)&\rightarrow&(1-G_\infty)e^{-t},\\
x^{(2)}(t)&\rightarrow&\frac{\lambda}{p_0},
\end{eqnarray}
leading to 
\begin{eqnarray}
x_0^+(t)&\rightarrow&x(0)c_2e^{-(1-\lambda)t},\\
x(t)&\rightarrow&x(0)c_2\frac{\lambda}{p_0}.
\end{eqnarray}
We do not reproduce the explicit expressions for the constants $c_1$ and $c_2$ as they are too long and not very illuminating. It suffices to say that in the case $\lambda=1/2$, $p_0=1/2$ used in subsection \ref{case5.3} it is $c_2=1/2$.

\section*{Acknowledgements}
Partial financial support has been received from the Agencia Estatal de Investigacion (AEI, Spain) and Fondo Europeo de Desarrollo Regional under Project PACSS RTI2018-093732-B-C21 (AEI/FEDER,UE) and the Spanish State Research Agency, through the Maria de Maeztu Program for units of Excellence in R\&D (MDM-2017-0711). A. F. P. acknowledges support by the Formacion de Profesorado Universitario (FPU14/00554) program of Ministerio de Educacion, Cultura y Deportes (MECD) (Spain). We thank J. F. Gracia for valuable discussions concerning the results of Ref. \cite{Juan}.

\section*{References}
\bibliographystyle{elsarticle-num-names}
\bibliography{aging-voter}

\begin{thebibliography}{9}
\providecommand{\natexlab}[1]{#1}
\providecommand{\url}[1]{\texttt{#1}}
\providecommand{\urlprefix}{URL }
\expandafter\ifx\csname urlstyle\endcsname\relax
  \providecommand{\doi}[1]{doi:\discretionary{}{}{}#1}\else
  \providecommand{\doi}[1]{doi:\discretionary{}{}{}\begingroup
  \urlstyle{rm}\url{#1}\endgroup}\fi
\providecommand{\bibinfo}[2]{#2}

\bibitem[{Pastor-Satorras and Vespignani(2001)}]{Disease1}
\bibinfo{author}{R.~Pastor-Satorras}, \bibinfo{author}{A.~Vespignani},
  \bibinfo{title}{Epidemic spreading in scale-free networks},
  \bibinfo{journal}{Physical review letters}
  \bibinfo{volume}{86}~(\bibinfo{number}{14}) (\bibinfo{year}{2001})
  \bibinfo{pages}{3200}.

\bibitem[{Pastor-Satorras et~al.(2015)Pastor-Satorras, Castellano, Van~Mieghem,
  and Vespignani}]{Disease2}
\bibinfo{author}{R.~Pastor-Satorras}, \bibinfo{author}{C.~Castellano},
  \bibinfo{author}{P.~Van~Mieghem}, \bibinfo{author}{A.~Vespignani},
  \bibinfo{title}{Epidemic processes in complex networks},
  \bibinfo{journal}{Reviews of modern physics}
  \bibinfo{volume}{87}~(\bibinfo{number}{3}) (\bibinfo{year}{2015})
  \bibinfo{pages}{925}.

\bibitem[{Abrams and Strogatz(2003)}]{Language1}
\bibinfo{author}{D.~M. Abrams}, \bibinfo{author}{S.~H. Strogatz},
  \bibinfo{title}{Linguistics: Modelling the dynamics of language death},
  \bibinfo{journal}{Nature} \bibinfo{volume}{424}~(\bibinfo{number}{6951})
  (\bibinfo{year}{2003}) \bibinfo{pages}{900}.

\bibitem[{Castell{\'o} et~al.(2006)Castell{\'o}, Egu{\'\i}luz, and
  San~Miguel}]{Language2}
\bibinfo{author}{X.~Castell{\'o}}, \bibinfo{author}{V.~M. Egu{\'\i}luz},
  \bibinfo{author}{M.~San~Miguel}, \bibinfo{title}{Ordering dynamics with two
  non-excluding options: bilingualism in language competition},
  \bibinfo{journal}{New Journal of Physics}
  \bibinfo{volume}{8}~(\bibinfo{number}{12}) (\bibinfo{year}{2006})
  \bibinfo{pages}{308}.

\bibitem[{Stauffer et~al.(2007)Stauffer, Castell{\'o}, Eguiluz, and
  San~Miguel}]{Language3}
\bibinfo{author}{D.~Stauffer}, \bibinfo{author}{X.~Castell{\'o}},
  \bibinfo{author}{V.~M. Eguiluz}, \bibinfo{author}{M.~San~Miguel},
  \bibinfo{title}{Microscopic Abrams--Strogatz model of language competition},
  \bibinfo{journal}{Physica A: Statistical Mechanics and its Applications}
  \bibinfo{volume}{374}~(\bibinfo{number}{2}) (\bibinfo{year}{2007})
  \bibinfo{pages}{835--842}.

\bibitem[{Vazquez et~al.(2010)Vazquez, Castell\'o, and Miguel}]{Language4}
\bibinfo{author}{F.~Vazquez}, \bibinfo{author}{X.~Castell\'o},
  \bibinfo{author}{M.~S. Miguel}, \bibinfo{title}{Agent based models of
  language competition: Macroscopic descriptions and order-disorder
  transition}, \bibinfo{journal}{J. Stat. Mech. Theory Exp}
  \bibinfo{volume}{2010}.

\bibitem[{Kirman(1993)}]{Kirman}
\bibinfo{author}{A.~Kirman}, \bibinfo{title}{Ants, rationality, and
  recruitment}, \bibinfo{journal}{Quart. J. Econ} \bibinfo{volume}{108}
  (\bibinfo{year}{1993}) \bibinfo{pages}{137}.

\bibitem[{Alfarano et~al.(2005)Alfarano, Lux, and Wagner}]{Markets1}
\bibinfo{author}{S.~Alfarano}, \bibinfo{author}{T.~Lux},
  \bibinfo{author}{F.~Wagner}, \bibinfo{title}{Time Estimation of agent-based
  models: The case of an asymmetric Herding model}, \bibinfo{journal}{Comput.
  Econ} \bibinfo{volume}{26} (\bibinfo{year}{2005}) \bibinfo{pages}{19}.

\bibitem[{Alfarano et~al.(2008)Alfarano, Lux, and Wagner}]{Markets2}
\bibinfo{author}{S.~Alfarano}, \bibinfo{author}{T.~Lux},
  \bibinfo{author}{F.~Wagner}, \bibinfo{title}{Time variation of higher moments
  in a financial market with heterogeneous agents: An analytical approach},
  \bibinfo{journal}{Econ. Dyn. Control} \bibinfo{volume}{32}
  (\bibinfo{year}{2008}) \bibinfo{pages}{101}.

\bibitem[{Alfarano and Milakovi\'c(2009)}]{Markets3}
\bibinfo{author}{S.~Alfarano}, \bibinfo{author}{M.~Milakovi\'c},
  \bibinfo{title}{Network structure and N-dependence in agent-based herding
  models}, \bibinfo{journal}{Journal of Economic Dynamics and Control}
  \bibinfo{volume}{33} (\bibinfo{year}{2009}) \bibinfo{pages}{78}.

\bibitem[{Gontis and Kononovicius(2017)}]{Markets4}
\bibinfo{author}{V.~Gontis}, \bibinfo{author}{A.~Kononovicius},
  \bibinfo{title}{Spurious Memory in Non-Equilibrium Stochastic Models of
  Imitative Behavior}, \bibinfo{journal}{Entropy} \bibinfo{volume}{19}
  (\bibinfo{year}{2017}) \bibinfo{pages}{387}.

\bibitem[{Carro et~al.(2015)Carro, Toral, and {San Miguel}}]{Carro1}
\bibinfo{author}{A.~Carro}, \bibinfo{author}{R.~Toral},
  \bibinfo{author}{M.~{San Miguel}}, \bibinfo{title}{{Markets, Herding and
  Response to External Information}}, \bibinfo{journal}{PloS one}
  \bibinfo{volume}{10}~(\bibinfo{number}{7}) (\bibinfo{year}{2015})
  \bibinfo{pages}{e0133287}.

\bibitem[{Kononovicius and Ruseckas(2019)}]{KONONOVICIUS2019171}
\bibinfo{author}{A.~Kononovicius}, \bibinfo{author}{J.~Ruseckas},
  \bibinfo{title}{Order book model with herd behavior exhibiting long-range
  memory}, \bibinfo{journal}{Physica A: Statistical Mechanics and its
  Applications} \bibinfo{volume}{525} (\bibinfo{year}{2019})
  \bibinfo{pages}{171 -- 191}.

\bibitem[{Barrat et~al.(2008)Barrat, Barthelemy, and Vespignani}]{babave08}
\bibinfo{author}{A.~Barrat}, \bibinfo{author}{M.~Barthelemy},
  \bibinfo{author}{A.~Vespignani}, \bibinfo{title}{Dynamical processes on
  complex networks}, \bibinfo{publisher}{Cambridge university press},
  \bibinfo{year}{2008}.

\bibitem[{Castellano et~al.(2009)Castellano, Fortunato, and Loreto}]{cafolo09}
\bibinfo{author}{C.~Castellano}, \bibinfo{author}{S.~Fortunato},
  \bibinfo{author}{V.~Loreto}, \bibinfo{title}{Statistical physics of social
  dynamics}, \bibinfo{journal}{Reviews of modern physics}
  \bibinfo{volume}{81}~(\bibinfo{number}{2}) (\bibinfo{year}{2009})
  \bibinfo{pages}{591}.

\bibitem[{Fern\'andez-Gracia et~al.(2014)Fern\'andez-Gracia, Suchecki, Ramasco,
  San~Miguel, and Egu\'{\i}luz}]{PhysRevLett.112.158701}
\bibinfo{author}{J.~Fern\'andez-Gracia}, \bibinfo{author}{K.~Suchecki},
  \bibinfo{author}{J.~J. Ramasco}, \bibinfo{author}{M.~San~Miguel},
  \bibinfo{author}{V.~M. Egu\'{\i}luz}, \bibinfo{title}{Is the Voter Model a
  Model for Voters?}, \bibinfo{journal}{Phys. Rev. Lett.} \bibinfo{volume}{112}
  (\bibinfo{year}{2014}) \bibinfo{pages}{158701}.

\bibitem[{Clifford and Sudbury(1973)}]{clsu73}
\bibinfo{author}{P.~Clifford}, \bibinfo{author}{A.~Sudbury}, \bibinfo{title}{A
  model for spatial conflict}, \bibinfo{journal}{Biometrika}
  \bibinfo{volume}{60}~(\bibinfo{number}{3}) (\bibinfo{year}{1973})
  \bibinfo{pages}{581--588}.

\bibitem[{Holley and Liggett(1975)}]{holi75}
\bibinfo{author}{R.~A. Holley}, \bibinfo{author}{T.~M. Liggett},
  \bibinfo{title}{Ergodic theorems for weakly interacting infinite systems and
  the voter model}, \bibinfo{journal}{The annals of probability}
  \bibinfo{volume}{3}~(\bibinfo{number}{4}) (\bibinfo{year}{1975})
  \bibinfo{pages}{643--663}.

\bibitem[{Vazquez and Egu{\'{\i}}luz(2008)}]{Vazquez_2008}
\bibinfo{author}{F.~Vazquez}, \bibinfo{author}{V.~M. Egu{\'{\i}}luz},
  \bibinfo{title}{Analytical solution of the voter model on uncorrelated
  networks}, \bibinfo{journal}{New Journal of Physics}
  \bibinfo{volume}{10}~(\bibinfo{number}{6}) (\bibinfo{year}{2008})
  \bibinfo{pages}{063011}.

\bibitem[{{Redner}(2018)}]{2018arXiv181111888R}
\bibinfo{author}{S.~{Redner}}, \bibinfo{title}{{Reality Inspired Voter Models:
  A Mini-Review}}, \bibinfo{journal}{arXiv e-prints}  (\bibinfo{year}{2018})
  \bibinfo{pages}{arXiv:1811.11888}.

\bibitem[{Krapivsky et~al.(2010)Krapivsky, Redner, and Ben-Naim}]{krrebe10}
\bibinfo{author}{P.~L. Krapivsky}, \bibinfo{author}{S.~Redner},
  \bibinfo{author}{E.~Ben-Naim}, \bibinfo{title}{A kinetic view of statistical
  physics}, \bibinfo{publisher}{Cambridge University Press},
  \bibinfo{year}{2010}.

\bibitem[{Liggett(2012)}]{li12}
\bibinfo{author}{T.~M. Liggett}, \bibinfo{title}{Interacting particle systems},
  vol. \bibinfo{volume}{276}, \bibinfo{publisher}{Springer Science \& Business
  Media}, \bibinfo{year}{2012}.

\bibitem[{Peralta et~al.(2018{\natexlab{a}})Peralta, Carro, Miguel, and
  Toral}]{Peralta_pair}
\bibinfo{author}{A.~F. Peralta}, \bibinfo{author}{A.~Carro},
  \bibinfo{author}{M.~S. Miguel}, \bibinfo{author}{R.~Toral},
  \bibinfo{title}{Stochastic pair approximation treatment of the noisy voter
  model}, \bibinfo{journal}{New Journal of Physics} \bibinfo{volume}{20}
  (\bibinfo{year}{2018}{\natexlab{a}}) \bibinfo{pages}{103045}.

\bibitem[{Carro et~al.(2016)Carro, Toral, and Miguel}]{Carro2}
\bibinfo{author}{A.~Carro}, \bibinfo{author}{R.~Toral}, \bibinfo{author}{M.~S.
  Miguel}, \bibinfo{title}{The noisy voter model on complex networks},
  \bibinfo{journal}{Scientific Reports} \bibinfo{volume}{6}
  (\bibinfo{year}{2016}) \bibinfo{pages}{24775}.

\bibitem[{Nyczka et~al.(2012)Nyczka, Sznajd-Weron, and Cis\l{}o}]{Nyczka2012}
\bibinfo{author}{P.~Nyczka}, \bibinfo{author}{K.~Sznajd-Weron},
  \bibinfo{author}{J.~Cis\l{}o}, \bibinfo{title}{Phase transitions in the
  $q$-voter model with two types of stochastic driving},
  \bibinfo{journal}{Phys. Rev. E} \bibinfo{volume}{86} (\bibinfo{year}{2012})
  \bibinfo{pages}{011105}.

\bibitem[{Nyczka and Sznajd-Weron(2013)}]{Nyczka2013}
\bibinfo{author}{P.~Nyczka}, \bibinfo{author}{K.~Sznajd-Weron},
  \bibinfo{title}{Anticonformity or Independence?---Insights from Statistical
  Physics}, \bibinfo{journal}{Journal of Statistical Physics}
  \bibinfo{volume}{151}~(\bibinfo{number}{1}) (\bibinfo{year}{2013})
  \bibinfo{pages}{174--202}.

\bibitem[{J\polhk{e}drzejewski(2017)}]{Jedrzejewski2017}
\bibinfo{author}{A.~J\polhk{e}drzejewski}, \bibinfo{title}{Pair approximation
  for the $q$-voter model with independence on complex networks},
  \bibinfo{journal}{Phys. Rev. E} \bibinfo{volume}{95} (\bibinfo{year}{2017})
  \bibinfo{pages}{012307}.

\bibitem[{Peralta et~al.(2018{\natexlab{b}})Peralta, Carro, Miguel, and
  Toral}]{Peralta}
\bibinfo{author}{A.~F. Peralta}, \bibinfo{author}{A.~Carro},
  \bibinfo{author}{M.~S. Miguel}, \bibinfo{author}{R.~Toral},
  \bibinfo{title}{Analytical and numerical study of the non-linear noisy voter
  on complex networks}, \bibinfo{journal}{Chaos} \bibinfo{volume}{28}
  (\bibinfo{year}{2018}{\natexlab{b}}) \bibinfo{pages}{075516}.

\bibitem[{Vieira and Anteneodo(2018)}]{PhysRevE.97.052106}
\bibinfo{author}{A.~R. Vieira}, \bibinfo{author}{C.~Anteneodo},
  \bibinfo{title}{Threshold $q$-voter model}, \bibinfo{journal}{Phys. Rev. E}
  \bibinfo{volume}{97} (\bibinfo{year}{2018}) \bibinfo{pages}{052106}.

\bibitem[{Raducha et~al.(2018)Raducha, Min, and Miguel}]{Raducha_2018}
\bibinfo{author}{T.~Raducha}, \bibinfo{author}{B.~Min}, \bibinfo{author}{M.~S.
  Miguel}, \bibinfo{title}{Coevolving nonlinear voter model with triadic
  closure}, \bibinfo{journal}{{EPL} (Europhysics Letters)}
  \bibinfo{volume}{124}~(\bibinfo{number}{3}) (\bibinfo{year}{2018})
  \bibinfo{pages}{30001}.

\bibitem[{Min and Miguel(2019)}]{Min_2019}
\bibinfo{author}{B.~Min}, \bibinfo{author}{M.~S. Miguel},
  \bibinfo{title}{Multilayer coevolution dynamics of the nonlinear voter
  model}, \bibinfo{journal}{New Journal of Physics}
  \bibinfo{volume}{21}~(\bibinfo{number}{3}) (\bibinfo{year}{2019})
  \bibinfo{pages}{035004}.

\bibitem[{Herrer{\'{i}}as-Azcu{\'{e}} and Galla(2019)}]{herrerias:2019}
\bibinfo{author}{F.~Herrer{\'{i}}as-Azcu{\'{e}}}, \bibinfo{author}{T.~Galla},
  \bibinfo{title}{{Consensus and diversity in multi-state noisy voter models}},
  \bibinfo{journal}{arXiv e-prints}  (\bibinfo{year}{2019})
  \bibinfo{pages}{arXiv:1903.09198}.

\bibitem[{Vazquez et~al.(2019)Vazquez, Loscar, and Baglietto}]{vazquez:2019}
\bibinfo{author}{F.~Vazquez}, \bibinfo{author}{E.~S. Loscar},
  \bibinfo{author}{G.~Baglietto}, \bibinfo{title}{{A multi-state voter model
  with imperfect copying}}, \bibinfo{journal}{arXiv e-prints}  (\bibinfo{year}{2019})
  \bibinfo{pages}{arXiv:1902.07253}.

\bibitem[{Khalil et~al.(2018)Khalil, Miguel, and Toral}]{Nagi}
\bibinfo{author}{N.~Khalil}, \bibinfo{author}{M.~S. Miguel},
  \bibinfo{author}{R.~Toral}, \bibinfo{title}{{Zealots in the mean-field noisy
  voter model}}, \bibinfo{journal}{Physical Review} \bibinfo{volume}{97}.

\bibitem[{Khalil and Toral(2019)}]{Nagi2}
\bibinfo{author}{N.~Khalil}, \bibinfo{author}{R.~Toral}, \bibinfo{title}{{The
  noisy voter model under the influence of contrarians}},
  \bibinfo{journal}{Physica A} \bibinfo{volume}{515} (\bibinfo{year}{2019})
  \bibinfo{pages}{81--92}.

\bibitem[{J\polhk{e}drzejewski and Sznajd-Weron(2018)}]{JEDRZEJEWSKI2018306}
\bibinfo{author}{A.~J\polhk{e}drzejewski}, \bibinfo{author}{K.~Sznajd-Weron},
  \bibinfo{title}{Impact of memory on opinion dynamics},
  \bibinfo{journal}{Physica A: Statistical Mechanics and its Applications}
  \bibinfo{volume}{505} (\bibinfo{year}{2018}) \bibinfo{pages}{306 -- 315}.

\bibitem[{Min et~al.(2013)Min, i.~Goh, and m.~Kim}]{Min1}
\bibinfo{author}{B.~Min}, \bibinfo{author}{K.~i.~Goh},
  \bibinfo{author}{I.~m.~Kim}, \bibinfo{title}{Suppression of epidemic
  outbreaks with heavy-tailed contact dynamics}, \bibinfo{journal}{Europhys.
  Lett} \bibinfo{volume}{103} (\bibinfo{year}{2013}) \bibinfo{pages}{50002}.

\bibitem[{Bogu{\~{n}}{\'{a}} et~al.(2014)Bogu{\~{n}}{\'{a}}, Lafuerza, Toral,
  and Serrano}]{Boguna2014}
\bibinfo{author}{M.~Bogu{\~{n}}{\'{a}}}, \bibinfo{author}{L.~Lafuerza},
  \bibinfo{author}{R.~Toral}, \bibinfo{author}{M.~{\'{A}}. Serrano},
  \bibinfo{title}{Simulating non-Markovian stochastic processes},
  \bibinfo{journal}{Physical Review E} \bibinfo{volume}{90}
  (\bibinfo{year}{2014}) \bibinfo{pages}{042108}.

\bibitem[{Starnini et~al.(2017)Starnini, Gleeson, and Bogu{\~n}\'a}]{Gleeson}
\bibinfo{author}{M.~Starnini}, \bibinfo{author}{J.~P. Gleeson},
  \bibinfo{author}{M.~Bogu{\~n}\'a}, \bibinfo{title}{Equivalence between
  Non-Markovian and {M}arkovian Dynamics in Epidemic Spreading Processes},
  \bibinfo{journal}{Phys. Rev. Lett} \bibinfo{volume}{118}
  (\bibinfo{year}{2017}) \bibinfo{pages}{128301}.

\bibitem[{Blythe and Anderson(1988)}]{Blythe}
\bibinfo{author}{S.~P. Blythe}, \bibinfo{author}{R.~M. Anderson},
  \bibinfo{title}{Variable Infectiousness in HIV Transmission Models},
  \bibinfo{journal}{Math. Med. Biol} \bibinfo{volume}{5} (\bibinfo{year}{1988})
  \bibinfo{pages}{181}.

\bibitem[{van Kampen(1998)}]{nonMarkov1}
\bibinfo{author}{N.~G. van Kampen}, \bibinfo{title}{Remarks on Non-Markov
  Processes}, \bibinfo{journal}{Braz. J. Phys} \bibinfo{volume}{28}
  (\bibinfo{year}{1998}) \bibinfo{pages}{90}.

\bibitem[{\L{}ucza(2005)}]{nonMarkov2}
\bibinfo{author}{J.~\L{}ucza}, \bibinfo{title}{Non-markovian stochastic
  processes: Colored noise}, \bibinfo{journal}{Chaos} \bibinfo{volume}{15}
  (\bibinfo{year}{2005}) \bibinfo{pages}{026107}.

\bibitem[{Fern{\'a}ndez-Gracia et~al.(2011)Fern{\'a}ndez-Gracia, Egu{\'\i}luz,
  and San~Miguel}]{Juan}
\bibinfo{author}{J.~Fern{\'a}ndez-Gracia}, \bibinfo{author}{V.~M.
  Egu{\'\i}luz}, \bibinfo{author}{M.~San~Miguel}, \bibinfo{title}{Update rules
  and interevent time distributions: Slow ordering versus no ordering in the
  voter model}, \bibinfo{journal}{Physical Review E}
  \bibinfo{volume}{84}~(\bibinfo{number}{1}) (\bibinfo{year}{2011})
  \bibinfo{pages}{015103}.

\bibitem[{Stark et~al.(2008)Stark, Tessone, and Schweitzer}]{Schweitzer}
\bibinfo{author}{H.-U. Stark}, \bibinfo{author}{C.~J. Tessone},
  \bibinfo{author}{F.~Schweitzer}, \bibinfo{title}{Decelerating Microdynamics
  Can Accelerate Macrodynamics in the Voter Model}, \bibinfo{journal}{Phys.
  Rev. Lett} \bibinfo{volume}{101} (\bibinfo{year}{2008})
  \bibinfo{pages}{018701}.

\bibitem[{Ben-Naim et~al.(1996)Ben-Naim, Frachebourg, and Krapivsky}]{befrkr96}
\bibinfo{author}{E.~Ben-Naim}, \bibinfo{author}{L.~Frachebourg},
  \bibinfo{author}{P.~L. Krapivsky}, \bibinfo{title}{Coarsening and persistence
  in the voter model}, \bibinfo{journal}{Physical Review E}
  \bibinfo{volume}{53}~(\bibinfo{number}{4}) (\bibinfo{year}{1996})
  \bibinfo{pages}{3078}.

\bibitem[{Sood and Redner(2005)}]{sore05}
\bibinfo{author}{V.~Sood}, \bibinfo{author}{S.~Redner}, \bibinfo{title}{Voter
  model on heterogeneous graphs}, \bibinfo{journal}{Physical review letters}
  \bibinfo{volume}{94}~(\bibinfo{number}{17}) (\bibinfo{year}{2005})
  \bibinfo{pages}{178701}.

\bibitem[{Suchecki et~al.(2005)Suchecki, Egu{\'\i}luz, and
  San~Miguel}]{suegsa05}
\bibinfo{author}{K.~Suchecki}, \bibinfo{author}{V.~M. Egu{\'\i}luz},
  \bibinfo{author}{M.~San~Miguel}, \bibinfo{title}{Voter model dynamics in
  complex networks: Role of dimensionality, disorder, and degree distribution},
  \bibinfo{journal}{Physical Review E}
  \bibinfo{volume}{72}~(\bibinfo{number}{3}) (\bibinfo{year}{2005})
  \bibinfo{pages}{036132}.

\bibitem[{Martins and Galam(2013)}]{martins:2013}
\bibinfo{author}{A.~C.~R. Martins}, \bibinfo{author}{S.~Galam},
  \bibinfo{title}{Building up of individual inflexibility in opinion dynamics},
  \bibinfo{journal}{Physical Review E} \bibinfo{volume}{87}
  (\bibinfo{year}{2013}) \bibinfo{pages}{042807}.

\bibitem[{Artime et~al.(2018)Artime, Peralta, Toral, Ramasco, and
  Miguel}]{Oriol}
\bibinfo{author}{O.~Artime}, \bibinfo{author}{A.~F. Peralta},
  \bibinfo{author}{R.~Toral}, \bibinfo{author}{J.~J. Ramasco},
  \bibinfo{author}{M.~S. Miguel}, \bibinfo{title}{Aging-induced phase
  transition}, \bibinfo{journal}{Phys. Rev. E} \bibinfo{volume}{98}
  (\bibinfo{year}{2018}) \bibinfo{pages}{032104}.

\bibitem[{{Artime} et~al.(2018){Artime}, {Carro}, {Peralta}, {Ramasco}, {San
  Miguel}, and {Toral}}]{2018arXiv181205378A}
\bibinfo{author}{O.~{Artime}}, \bibinfo{author}{A.~{Carro}},
  \bibinfo{author}{A.~F. {Peralta}}, \bibinfo{author}{J.~J. {Ramasco}},
  \bibinfo{author}{M.~{San Miguel}}, \bibinfo{author}{R.~{Toral}},
  \bibinfo{title}{{Herding and idiosyncratic choices: Nonlinearity and
  aging-induced transitions in the noisy voter model}}, \bibinfo{journal}{arXiv
  e-prints}  \bibinfo{eid}{arXiv:1812.05378}.

\bibitem[{Considine et~al.(1989)Considine, Redner, and
  Takayasu}]{PhysRevLett.63.2857}
\bibinfo{author}{D.~Considine}, \bibinfo{author}{S.~Redner},
  \bibinfo{author}{H.~Takayasu}, \bibinfo{title}{Comment on ``Noise-induced
  bistability in a Monte Carlo surface-reaction model''},
  \bibinfo{journal}{Phys. Rev. Lett.} \bibinfo{volume}{63}
  (\bibinfo{year}{1989}) \bibinfo{pages}{2857--2857}.

\bibitem[{Granovsky and Madras(1995)}]{Granovsky}
\bibinfo{author}{B.~L. Granovsky}, \bibinfo{author}{N.~Madras},
  \bibinfo{title}{The noisy voter model}, \bibinfo{journal}{Stoch. Proc. Appl.}
  \bibinfo{volume}{55} (\bibinfo{year}{1995}) \bibinfo{pages}{23}.

\bibitem[{Wu et~al.(2010)Wu, Zhou, Xiao, Kurths, and Schellnhuber}]{Wu}
\bibinfo{author}{Y.~Wu}, \bibinfo{author}{C.~Zhou}, \bibinfo{author}{J.~Xiao},
  \bibinfo{author}{J.~Kurths}, \bibinfo{author}{H.~J. Schellnhuber},
  \bibinfo{title}{Evidence for a bimodal distribution in human communication},
  \bibinfo{journal}{PNAS} \bibinfo{volume}{107} (\bibinfo{year}{2010})
  \bibinfo{pages}{18803}.

\bibitem[{Candia et~al.(2008)Candia, Gonz\'alez, Wang, Schoenharl, Madey, and
  Barab\'asi}]{Candia}
\bibinfo{author}{J.~Candia}, \bibinfo{author}{M.~C. Gonz\'alez},
  \bibinfo{author}{P.~Wang}, \bibinfo{author}{T.~Schoenharl},
  \bibinfo{author}{G.~Madey}, \bibinfo{author}{A.~Barab\'asi},
  \bibinfo{title}{Uncovering individual and collective human dynamics from
  mobile phone records}, \bibinfo{journal}{J. Phys. A: Math. Theor}
  \bibinfo{volume}{41} (\bibinfo{year}{2008}) \bibinfo{pages}{224015}.

\bibitem[{P\'erez et~al.(2016)P\'erez, Klemm, and Egu\'iluz}]{Perez}
\bibinfo{author}{T.~P\'erez}, \bibinfo{author}{K.~Klemm},
  \bibinfo{author}{V.~M. Egu\'iluz}, \bibinfo{title}{Competition in the
  presence of aging: dominance, coexistence, and alternation between states},
  \bibinfo{journal}{Sci. Rep} \bibinfo{volume}{6} (\bibinfo{year}{2016})
  \bibinfo{pages}{21128}.

\bibitem[{Artime et~al.(2017{\natexlab{a}})Artime, Ramasco, and
  Miguel}]{Oriol3}
\bibinfo{author}{O.~Artime}, \bibinfo{author}{J.~J. Ramasco},
  \bibinfo{author}{M.~S. Miguel}, \bibinfo{title}{Dynamics on networks:
  competition of temporal and topological correlations}, \bibinfo{journal}{Sci.
  Rep} \bibinfo{volume}{7} (\bibinfo{year}{2017}{\natexlab{a}})
  \bibinfo{pages}{41627}.

\bibitem[{Artime et~al.(2017{\natexlab{b}})Artime, Gracia, Ramasco, and
  Miguel}]{Oriol2}
\bibinfo{author}{O.~Artime}, \bibinfo{author}{J.~F. Gracia},
  \bibinfo{author}{J.~J. Ramasco}, \bibinfo{author}{M.~S. Miguel},
  \bibinfo{title}{Joint effect of ageing and multilayer structure prevents
  ordering in the voter model}, \bibinfo{journal}{Sci. Rep} \bibinfo{volume}{7}
  (\bibinfo{year}{2017}{\natexlab{b}}) \bibinfo{pages}{7166}.

\bibitem[{Escaff et~al.(2018)Escaff, Toral, {Van Den Broeck}, and
  Lindenberg}]{Escaff:2018}
\bibinfo{author}{D.~Escaff}, \bibinfo{author}{R.~Toral},
  \bibinfo{author}{C.~{Van Den Broeck}}, \bibinfo{author}{K.~Lindenberg},
  \bibinfo{title}{{A continuous-time persistent random walk model for
  flocking}}, \bibinfo{journal}{Chaos}
  \bibinfo{volume}{28}~(\bibinfo{number}{7}) (\bibinfo{year}{2018})
  \bibinfo{pages}{075507}.

\bibitem[{Artime et~al.(2018)Artime, Khalil, Toral, and Miguel}]{Nagi3}
\bibinfo{author}{O.~Artime}, \bibinfo{author}{N.~Khalil},
  \bibinfo{author}{R.~Toral}, \bibinfo{author}{M.~S. Miguel},
  \bibinfo{title}{{First-passage distributions for the one-dimensional
  Fokker-Planck equation}}, \bibinfo{journal}{Physical Review}
  \bibinfo{volume}{98}.

\bibitem[{van Kampen(2007)}]{vanKampen:2007}
\bibinfo{author}{N.~van Kampen}, \bibinfo{title}{{Stochastic Processes in
  Physics and Chemistry}}, \bibinfo{publisher}{North-Holland},
  \bibinfo{address}{Amsterdam}, \bibinfo{edition}{third} edn.,
  \bibinfo{year}{2007}.

\bibitem[{Toral and Colet(2014)}]{Toral-Colet:2014}
\bibinfo{author}{R.~Toral}, \bibinfo{author}{P.~Colet},
  \bibinfo{title}{{Stochastic Numerical Methods: An Introduction for Students
  and Scientists}}, \bibinfo{publisher}{Wiley},
  \urlprefix\url{http://eu.wiley.com/WileyCDA/WileyTitle/productCd-3527411496.html},
  \bibinfo{year}{2014}.

\bibitem[{Peralta and Toral(2018)}]{Peralta_moments}
\bibinfo{author}{A.~F. Peralta}, \bibinfo{author}{R.~Toral},
  \bibinfo{title}{System-size expansion of the moments of a master equation},
  \bibinfo{journal}{Chaos} \bibinfo{volume}{28} (\bibinfo{year}{2018})
  \bibinfo{pages}{106303}.

\bibitem[{Ozaita(2018)}]{Ozaita}
\bibinfo{author}{J.~Ozaita}, \bibinfo{title}{Noisy voter model with partial
  aging and anti-aging}, Master's thesis, \bibinfo{school}{University of the
  Balearic Islands, Palma, Spain}, \bibinfo{year}{2018}.

\bibitem[{Day(1967)}]{JamesThomas}
\bibinfo{author}{J.~T. Day}, \bibinfo{title}{{Note on the Numerical Solution of
  Integro-Differential Equations}}, \bibinfo{journal}{The Computer Journal}
  \bibinfo{volume}{9}~(\bibinfo{number}{4}) (\bibinfo{year}{1967})
  \bibinfo{pages}{394--395}.

\bibitem[{qpo(2019)}]{qpochhammer}
\urlprefix\url{https://en.wikipedia.org/wiki/Q-Pochhammer\_symbol},
  \bibinfo{year}{2019}.

\bibitem[{Note1(????)}]{Note1}
\bibinfo{note}{A similar qualitative behavior is observed for the
  magnetization. In fact, for the all-to-all connectivity considered later in
  this paper, one has simply $\rho (t)=\protect \frac 12\left [1-m(t)^2\right
  ]$, although in other networks $\rho (t)$ must be considered as an
  independent variable.}

\bibitem[{Note2(????)}]{Note2}
\bibinfo{note}{Strictly speaking, the aforementioned power-law
  dependence was reported in \cite {Juan} for the cumulative inter-event time
  distribution $C(t)$. One of the authors of that reference (J. F. Gracia,
  private communication) has confirmed to us that the aforementioned asymptotic
  time dependence of $\rho (t)$ was also observed, but not displayed.}

\bibitem[{Note3(????)}]{Note3}
\bibinfo{note}{Simpler expressions are obtained if $c$ is an integer
  number, e.g. $f(x)=\left [p_\infty (1-x)\right ]^{-\protect \frac
  {1-p0(1-x)}{1-p_\infty (1-x)}}$, if $c=1$.}

\bibitem[{Note4(????)}]{Note4}
\bibinfo{note}{In some cases this limit is non-trivial and must be
  performed carefully. Specially when $p_{i}$ decays very fast as in the
  exponential example. A possibility is to consider $M(t)$ as time dependent
  and a phenomenological reasoning based on imposing convergence of $f(x) \sim
  M(t) \DOTSB \prod _{k=0}^{t} \alpha_{k}(1-x)$ with $t\rightarrow
  \infty $ suggests $M(t)\sim \left [ \DOTSB \prod_{k=0}^{t} (1-p_{k}
  x(0)) \right ]^{-1}$. This is similar to imposing $\protect \frac {d
  x_{i}^\pm }{d t}=0$ only for $i < t$ lower than the simulation time $t$.}

\bibitem[{Note5(????)}]{Note5}
\bibinfo{note}{A possibility is to use the Laplace transform in Eqs.
  (\ref {order3},\ref {order4}).}

\bibitem[{Note6(????)}]{Note6}
\bibinfo{note}{Note here that $x_{0}^{+}(\infty ) = p x(0)$ in apparent
  contradiction to the result of section \ref {sec:master}, $x_{0,\protect
  \mathrm {st}}^+=px_{\protect \mathrm {st}}(1-x_{\protect \mathrm {st}})$. But
  this is a product of the linearization in Eqs. (\ref {order3},\ref
  {order4}).}

\bibitem[{Note7(????)}]{Note7}
\bibinfo{note}{It is possible to try a power-series expansion
  $x(t)=\DOTSB \sum_{k=0}^\infty a_kt^k$ and $x_0^+(t)=\DOTSB \sum_{k=0}^\infty b_kt^k$ and obtain recurrence relations for the
  coefficients $a_k,\protect \tmspace +\thinmuskip {.1667em}b_k$, but we have
  not been able to sum analytically the resulting series.}

\end{thebibliography}

\end{document}